\documentclass[twocolumn]{aastex62}
\pdfoutput=1

\usepackage{graphicx}
\usepackage{journals}
\usepackage[symbol]{footmisc}
\usepackage{url}
\usepackage{color}
\usepackage{multirow}
\usepackage{xspace}
\usepackage{amsmath}
\usepackage{mathtools}

\newcommand{\ergscm}{\ensuremath{{\rm erg\,s^{-1}\,cm^{-2}}}\xspace}

\graphicspath{{./}{figures/}}

\received{April 26, 2018}
\revised{--}
\accepted{--}
\submitjournal{ApJ}

\shorttitle{IMBH population identified as AGN}
\shortauthors{Chilingarian et al.}

\begin{document}

\title{A Population of \textit{Bona Fide} Intermediate Mass Black Holes Identified as Low Luminosity Active Galactic Nuclei}

\correspondingauthor{Igor Chilingarian}
\email{igor.chilingarian@cfa.harvard.edu, chil@sai.msu.ru}

\author[0000-0002-7924-3253]{Igor V. Chilingarian}
\affiliation{Smithsonian Astrophysical Observatory, 60 Garden St. MS09, Cambridge, MA 02138, USA}
\affiliation{Sternberg Astronomical Institute, M.V.Lomonosov Moscow State University, Universitetsky prospect 13, Moscow, 119992, Russia}

\author{Ivan Yu. Katkov}
\affiliation{Sternberg Astronomical Institute, M.V.Lomonosov Moscow State University, Universitetsky prospect 13, Moscow, 119992, Russia}

\author{Ivan Yu. Zolotukhin}
\affiliation{Sternberg Astronomical Institute, M.V.Lomonosov Moscow State University, Universitetsky prospect 13, Moscow, 119992, Russia}
\affiliation{Special Astrophysical Observatory of the Russian Academy of Sciences, Nizhnij Arkhyz 369167, Russia}
\affiliation{Universit\'e de Toulouse; UPS-OMP, IRAP, 9 avenue du Colonel Roche, BP 44346, F-31028 Toulouse Cedex 4, France}

\author{Kirill A. Grishin}
\affiliation{Sternberg Astronomical Institute, M.V.Lomonosov Moscow State University, Universitetsky prospect 13, Moscow, 119992, Russia}
\affiliation{Department of Physics, M.V. Lomonosov Moscow State University, 1 Vorobyovy Gory, Moscow, 119991, Russia}

\author{Yuri Beletsky}
\affiliation{Las Campanas Observatory, Carnegie Institution of Washington, Colina el Pino, Casilla 601 La Serena, Chile}

\author{Konstantina Boutsia}
\affiliation{Las Campanas Observatory, Carnegie Institution of Washington, Colina el Pino, Casilla 601 La Serena, Chile}

\author{David J. Osip}
\affiliation{Las Campanas Observatory, Carnegie Institution of Washington, Colina el Pino, Casilla 601 La Serena, Chile}



\begin{abstract}
Nearly every massive galaxy harbors a supermassive black hole (SMBH) in its nucleus. SMBH masses are millions to billions $M_{\odot}$, and they correlate with properties of spheroids of their host galaxies. While the SMBH growth channels, mergers and gas accretion, are well established, their origin remains uncertain: they could have either emerged from massive ``seeds'' ($10^5-10^6 M_{\odot}$) formed by direct collapse of gas clouds in the early Universe or from smaller ($100 M_{\odot}$) black holes, end-products of first stars. The latter channel would leave behind numerous intermediate mass black holes (IMBHs, $10^2-10^5 M_{\odot}$). Although many IMBH candidates have been identified, none is accepted as definitive, thus their very existence is still debated. Using data mining in wide-field sky surveys and applying dedicated analysis to archival and follow-up optical spectra, we identified a sample of 305 IMBH candidates having masses $3\times10^4<M_{\mathrm{BH}}<2\times10^5 M_{\odot}$, which reside in galaxy centers and are accreting gas that creates characteristic signatures of a type~I active galactic nucleus (AGN). We confirmed the AGN nature of ten sources (including five previously known objects which validate our method) by detecting the X-ray emission from their accretion discs, thus defining the first \emph{bona fide} sample of IMBHs in galactic nuclei.  All IMBH host galaxies possess small bulges and sit on the low-mass extension of the $M_{\mathrm{BH}}-M_{\mathrm{bulge}}$ scaling relation suggesting that they must have experienced very few if any major mergers over their lifetime.  The very existence of nuclear IMBHs supports the stellar mass seed scenario of the massive black hole formation.
\end{abstract}



\section{Introduction and motivation}
\label{sec:intro}
The existence of stellar mass black holes \citep{LIGO1} and giant black holes millions to billions times more massive than the Sun is observationally established \citep{SgrA,miyoshi95}. Massive black holes in galaxy centers are believed to co-evolve with spheroids of their hosts \citep{kormendy13}, grow via coalescences during galaxy mergers \citep{2005LRR.....8....8M} and by accreting gas during the AGN/quasar phase \citep{2012Sci...337..544V}, however, their origin still remains unclear: they could have either emerged from massive ``seeds'' ($10^5-10^6 M_{\odot}$) formed by direct collapse of large gas clouds in the early Universe \citep{loeb94} or from smaller ($100 M_{\odot}$) black holes, end-products of first stars \citep{madau01}, which must have also created a rich, yet undetected population of intermediate mass black holes (IMBHs, $10^2-10^5 M_{\odot}$).

Two observational phenomena allow us to detect and estimate masses of central black holes in large samples of galaxies: (i) dynamic signatures observed as high rotational velocities or velocity dispersions of stars and gas in circumnuclear regions of galaxies \citep{miyoshi95,1995ARA&A..33..581K,seth14}; (ii) AGN and quasars, which appear when a massive black hole is caught while accreting gas \citep{elvis00} and, hence, growing. The discovery of quasars in the early Universe ($z > 6.3$, that is 750--900~Myr after the Big Bang) hosting super-massive black holes (SMBHs) as heavy as $10^{10} M_{\odot}$ \citep{mortlock11,2015Natur.518..512W} cannot be explained by gas accretion on stellar mass black hole seeds ($\lesssim 100 M_{\odot}$) alone. Even if formed right after the Big Bang by the first generation of \emph{Population}~{\sc iii} stars, it would take over 1~Gyr to foster an SMBH because the accretion rate cannot significantly exceed the Eddington limit during extended periods of time. \emph{Population}~{\sc iii} stars might have formed in dense clusters in primordial density fluctuations, which could then evolve into more massive SMBH seeds by collisions and/or core collapse \citep{2004Natur.428..724P}. Alternatively, the rapid inflow and subsequent direct collapse of gas clouds \citep{loeb94,begelman06} can form massive seeds ($M>10^5-10^6 M_{\odot}$). The latter scenario solves the SMBH early formation puzzle but leads to a gap in the present-day black hole mass function in the IMBH regime ($100\lesssim M_{\mathrm{BH}}\lesssim1\times10^5 M_{\odot}$), whereas stellar mass seeds should leave behind a large number of IMBHs. Therefore, the elusive IMBH population holds a clue to the understanding of SMBH formation.

The first evidence for the existence of IMBHs came in late 80s, when two dwarf galaxies with stellar masses of about $10^9 M_{\odot}$ hosting AGN were identified: \object{Pox~52}, a dwarf elliptical galaxy \citep{KSB87,BHRS04} and \object{NGC~4395}, a low-luminosity spiral \citep{FS89,WH06}. They both host central black holes with the mass estimates around $3\times10^5 M_{\odot}$ \citep{FH03,Peterson+05,Thornton+08} and are nowadays considered too massive to be called IMBH, however, they ignited the interest towards search for less massive examples.

Despite substantial observing time investments over the past two decades, only a few IMBH candidates were identified with reliable mass estimates: (i) the serendipitously discovered hyperluminous X-ray source \emph{HLX-1} in a nearby galaxy \citep{webb12} having the mass between $10^4$ and $10^5 M_{\odot}$ estimated from the X-ray flux and radio emission of the relativistic jet, which, however, may be a stellar mass black hole accreting in the super-critical regime with a beamed X-ray radiation along the line-of-sight \citep{king14}; (ii) a 4,000~$M_{\odot}$ IMBH in the globular cluster \emph{47~Tuc} detected using stellar dynamics and pulsar timing \citep{kiziltan17}, although one has to keep in mind that several past claims of IMBHs in globular clusters \citep{2010ApJ...719L..60N} were refuted by subsequent analysis \citep{2017MNRAS.468.4429Z}; (iii) several low luminosity AGN in dwarf galaxies found using optical spectra and later confirmed in X-ray \citep{dong07,reines13,baldassare15} ($5\times10^4<M_{\mathrm{BH}}<3\times10^5 M_{\odot}$) but the search approach excluded luminous galaxies and involved substantial amount of manual data analysis applied on a \emph{per object} basis. Here we call ``reliable'' the black hole mass estimates techniques, which have calibration uncertainties of at most a factor of 2--3, such as reverberation mapping, stellar dynamics, pulsar timing, X-ray variability, broad H$\alpha$ scaling. We do not consider IMBH candidates relying on some average Eddington ratios in AGN, $M_{BH} - \sigma_{\mathrm{bulge}}$ relation or the fundamental plane of the black hole activity, which have intrinsic uncertainties of 0.8--1.5~dex. 

In this paper we present the results of the first systematic search for IMBHs in AGN without applying {\it a priori} pre-selection filtering criteria to the input galaxy sample.
We developed an automated workflow that analyzed 1 million galaxies spectra from the SDSS Data Release 7 (DR7) spectroscopic catalog and measured central BH masses for those objects that demonstrate broad H$\alpha$ line and narrow-line photoionization signatures of accreting BHs.
Throughout this work we define an IMBH as an object in the mass range between $10^2$ and $10^5 M_{\odot}$ and we look for IMBH candidates having a mass $<2\times10^5 M_{\odot}$ because of the internal precision of the virial mass estimate of about $\sim0.3$~dex.

\section{Data analysis approach}

\subsection{An automated IMBH search workflow}

\begin{figure*}
\includegraphics[width=\hsize]{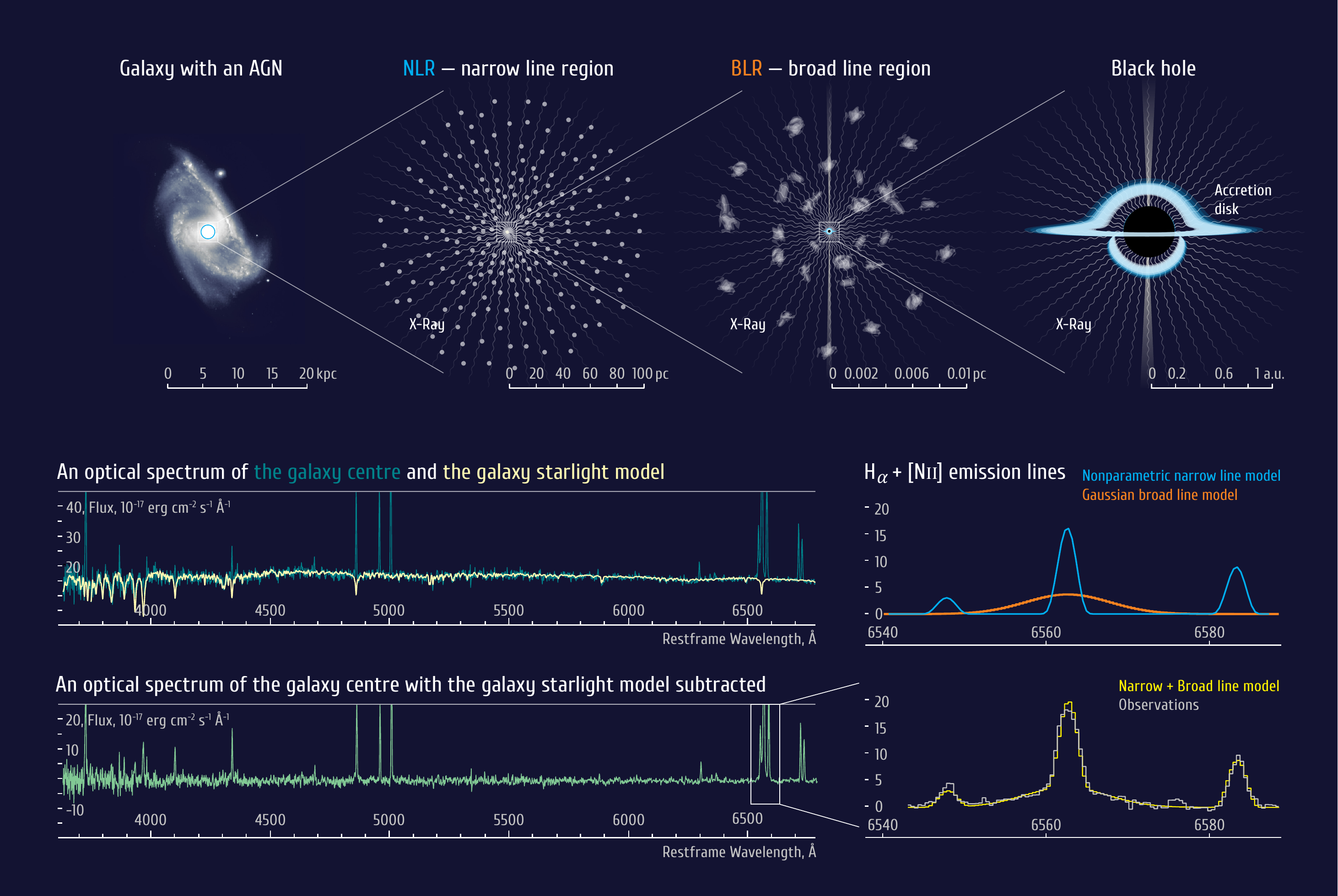}
\caption{A black hole mass determination in AGN from optical spectra. Top row: A black hole with an accretion disc ionizes the interstellar medium in its host galaxy. Dense gas clouds in the immediate vicinity of the black hole (0.001--0.1~pc; broad line region or BLR) are virialized and move at velocities up-to thousands km/s, thus broadening recombination lines originating from allowed transitions. Rarefied gas clouds further away from the black hole ($\lesssim 1$~kpc; narrow line region or NLR) move much slower (up-to hundreds km/s) and emit also in forbidden transitions, however, the narrow line shape depends on the exact NLR morphology. Middle and bottom rows: We model the stellar content of a galaxy by fitting its observed spectrum against a grid of stellar population models; then fit emission line residuals, first by using the same non-parametric shape for all detected lines and then by adding Gaussian broad line components in the hydrogen Balmer lines. If the fitting results differ significantly, we estimate the virial black hole mass from the broad line component width and luminosity using the calibration by \citet{reines13}.
\label{fig_spec}}
\end{figure*}

An AGN creates specific signatures in an optical spectrum of a galaxy \citep{elvis00}: X-ray photons from the corona of an accretion disc around the black hole ionize gas out to a few kiloparsec away, which then produces easily detectable emission lines (Fig.~\ref{fig_spec}). The width and the flux of an allowed recombination line (e.g. hydrogen H$\alpha$) emitted from the broad line region in the immediate vicinity of a central black hole provide a virial estimate of its mass \citep{greene05,reines13}. We have designed an automated workflow that uses data mining in optical and X-ray astronomical data archives publicly available in the international Virtual Observatory to search for AGN signatures of IMBHs.

The workflow automatically analyzed (Fig.~\ref{fig_spec}) about 1,000,000 optical spectra of galaxies and quasars from the legacy sample of the Sloan Digital Sky Survey Data Release 7 (SDSS DR7, \citealp{SDSS_DR7}) without any pre-selection on host galaxy luminosity or redshift. We used a non-parametric representation of a narrow emission line profile \citep{RCSED}, which produced lower fitting residuals in Balmer lines and allowed us to detect fainter broad line components, thus boosting the sensitivity of our analysis technique. Then we took the resulting sample of galaxies with broad line detections, computed emission line flux ratios [O{\sc iii}]/H$\beta$ and [N{\sc ii}]/H$\alpha$ from narrow-line components, and used the Baldwin-Phillips-Terlevich (BPT) \citep{BPT81} diagnostics to reject objects where the ionization was induced by star formation, because such objects often have broad Balmer lines originating from transient stellar events \citep{baldassare16} rather than from AGN. After that, we eliminated statistically insignificant measurements by filtering the sample on relative strengths and widths of narrow and broad line components, signal-to-noise ratios, relative radial velocity offsets, and assembled a list of 305 candidates in the IMBH mass range ($M_{\rm BH} < 2 \times 10^5 M_{\odot}$). 

This list (further referred as \textit{the parent sample}) included all known nuclear IMBH candidates \citep{dong07,reines13,baldassare15} except those which fell on the star forming sequence in the BPT diagram; our black hole mass estimates agreed within uncertainties with those obtained from dedicated deep spectroscopic observations \citep{reines13,baldassare15}. Then we searched in data archives of \emph{Chandra}, \emph{XMM-Newton}, and \emph{Swift} orbital X-ray observatories and detected 10 X-ray counterparts for candidates with virial masses as low as $4.3\times10^4 M_{\odot}$. Five of them were mentioned in the literature \citep{reines13,baldassare15} and one object had a spatially extended X-ray emission probably related to star formation which we excluded from further analysis. All 4 remaining X-ray point sources were observed serendipitously.

\subsection{Non-parametric emission line analysis}

We developed a dedicated technique to analyze optical spectra, which allows us to estimate a central black hole mass by measuring a broad component in allowed emission lines \citep{greene05}. As an input dataset for our study we use galaxy spectra from SDSS DR7, which we re-analyzed and presented in the value-added catalog RCSED \citep{RCSED}. We extended RCSED by adding AGN spectra classified as quasars in SDSS.
This input sample contains 938,487 galaxy spectra of 878,138 unique objects.
We analyzed the follow-up optical spectra obtained with the 6.5-m Magellan telescope in the same fashion. We fitted and subtracted stellar continuum in SDSS spectra using the {\sc NBursts} full spectrum fitting technique \citep{CPSA07,Chilingarian07a} and then measured emission lines.

The core of our analysis method is a simultaneous fitting of all strong emission lines (H$\beta$, [O{\sc iii}], [N{\sc ii}], H$\alpha$, [S{\sc II}]) by a linear combination of a narrow line component having a non-parametric shape and a broad Gaussian component in the Balmer lines.
The broad-line component parameters, velocity dispersion ($\sigma_{\rm BLR}$) and the central radial velocity of the BLR component are fitted in a non-linear minimization loop.
All other parameters are fitted linearly within it at every evaluation of the function using an iterative procedure that includes the following two steps: (i)
we determine fluxes of all emission line components solving a linear problem with the non-negative constrain; (ii) then we recover the shape of the narrow line component in a non-parametric way by solving a linear convolution problem with the regularization, which requires a smoothness of a solution.
By using a non-parametric NLR component shape, we can successfully model complex gas kinematics and avoid the degeneracy, which affects the traditionally used multiple Gaussian profile decomposition \citep{greene05,reines13}, because Gaussian functions are not orthogonal and, therefore, do not form a basis.
We then repeated our analysis by excluding a BLR component from the model in order to compare the $\chi^2$ values between the two approaches and conclude whether adding a BLR component improved the fitting quality in a statistically significant way.

%


\subsection{Constructing the sample: IMBH selection criteria}

Having obtained the flux and the width of the H$\alpha$ broad component in all 938,487 spectra of the input sample, we use the conservative empirical calibration to estimate a virial black hole mass \citep{reines13}:

\begin{equation}
\begin{array}{l}
M_{\rm BH} = 3.72 \times 10^6  \, ({\rm FWHM}_{\rm H\alpha}/10^3 {\rm \,km\,s^{-1}})^{2.06} \\
\hspace{40pt} \times (L_{\rm H\alpha}/10^{42} {\rm \,erg\,s^{-1}})^{0.47} \, M_{\odot}
\end{array}
\end{equation}

By comparing a broad H$\alpha$ based virial estimates with other black hole mass measurement techniques (e.g. reverberation, stellar dynamics) it was demonstrated \citep{dong12} that they agree within 0.3~dex, i.e. a factor of 2.
One, however, has to keep in mind that this uncertainty also includes statistical and systematic errors of black hole mass measurements used in the calibration, which might significantly contribute to the 0.3~dex error budget.
We use this as a rough estimate of the systematic uncertainty of our method, which also defines the mass range of our search: $M_{\rm BH}<2\times10^5 M_{\odot}$.

We then apply multiple selection criteria in order to filter reliable IMBH candidates from the input sample and eliminate spurious broad line detections:

\begin{itemize}
\item $M_{\rm BH} < 2 \times 10^5\, M_{\odot}$ in order to select objects in the IMBH mass range given the assumed 0.3~dex systematic uncertainty.
\item No night sky airglow lines falling in the regions around H$\alpha$+[N{\sc ii}], [O{\sc iii}] 5007\AA, and H$\beta$ 4861\AA, which we use for the spectral line profile fitting and decomposition.
\item Empirical constraint that the width of the broad line component is at least $\sqrt[]{5}$ times larger than that of the narrow line component.
\item Signal-to-noise ratio exceeding 3 in every emission line used in the BPT \citep{BPT81} classification (H$\beta$, [O{\sc iii}], [N{\sc ii}], H$\alpha$), which ensures its reliability.
\item The BPT classification is ``AGN'' or ``composite'' \citep{kewley06}, that discards star-forming galaxies because broad line components in them are often transient \citep{baldassare16}.
\item The H$\alpha$/H$\beta$ Balmer decrement for both narrow and broad line components $<4$.
\item Statistical error on $M_{\rm BH}$ better than 33\%
\item $|v_{\rm BLR}-v_{\rm NLR}|<3 \sigma_{\rm NLR}$ to reject strongly asymmetric BLR profiles.
\end{itemize}

These criteria joined together with boolean {\sc and} form our main selection filter.
It leaves a sample of 305 IMBH candidates out of nearly 1 million input spectra.
We call it {\it the parent sample}.

\section{Follow-up observations and analysis of new and archival data}

In order to exclude possible transient phenomena such as core collapse supernova or tidal disruption events, we followed up 3 galaxies with X-ray counterparts, 4 targets selected for our X-ray observations, and 5 additional IMBH candidates (12 targets in total) using the intermediate resolution Magellan Echellette Spectrograph (MagE) at the 6.5-m Magellan Baade telescope (see Table~\ref{obslog}). We processed MagE spectra through our data analysis technique and obtained independent second-epoch IMBH mass estimates consistent within uncertainties with SDSS (see Table~\ref{tableimbh}) for 8 galaxies. We did not detect a broad H$\alpha$ component in 3 objects. 

\begin{figure*}
\includegraphics[width=\hsize]{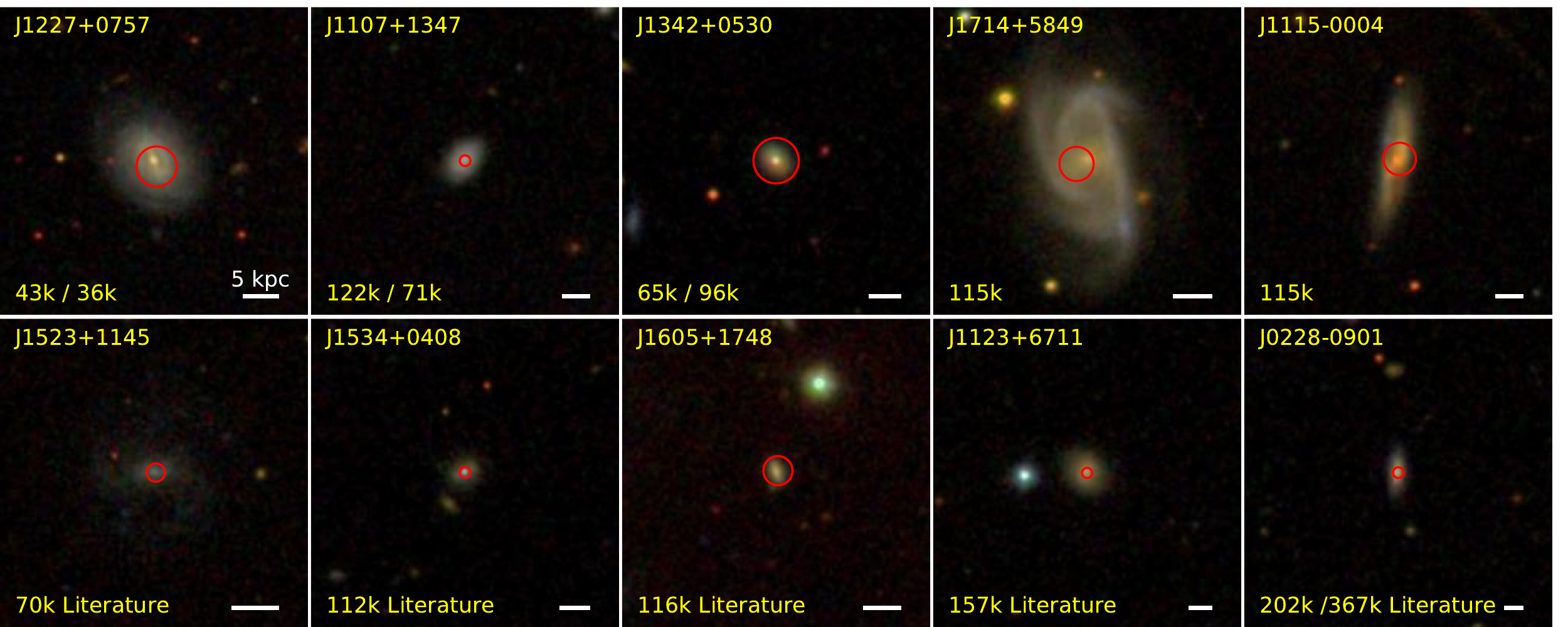}
\caption{Optical images of ten IMBH host galaxies.
Sloan Digital Sky Survey images of galaxies hosting IMBHs detected by our automated data analysis workflow demonstrate low luminosity spheroidal stellar systems or spiral galaxies with small bulges. The locations of X-ray counterparts with the corresponding 3$\sigma$ positional uncertainties is shown by red circles. The bottom row displays objects mentioned in the literature previously, which our workflow has successfully recovered. A virial mass estimate in $M_{\odot}$ from the analysis of SDSS spectra is shown in the bottom left corner of every panel followed by an estimate from MagE when available, the physical scale in the host galaxy plane is in the bottom right.
\label{fig_sample}}
\end{figure*}

The observed flux in the forbidden oxygen line [O{\sc iii}] ($\lambda=5007$~\AA) in AGN correlates with the X-ray luminosity $L_X$ \citep{2005ApJ...634..161H}, because the NLR is ionized by energetic photons originating from the active nucleus. Using this correlation, we selected 4 IMBH candidates with estimated X-ray fluxes $>5\times10^{-15}$~erg~cm$^{-2}$~s$^{-1}$ which can be detected in a 10,000~s exposure for follow-up X-ray observations. We obtained a solid confirmation for one source using \emph{Chandra} ($M_{\mathrm{BH}}=1.2\times10^5 M_{\odot}$, MagE) and a low-confidence detection for another source using \emph{XMM-Newton} ($M_{\mathrm{BH}}=7.5\times10^4 M_{\odot}$, SDSS, no broad H$\alpha$ component detected with MagE). The two remaining objects were not detected in X-ray suggesting that either we observed them in a low phase of activity and they fell below the [O{\sc iii}]--$L_X$ correlation or that the broad lines were due to transient phenomena. Finally, we ended up with a sample of 10 \emph{bona fide} broad-line AGN with virial black hole masses between 43,000 and 202,000 $M_{\odot}$ estimated from SDSS spectra having point source X-ray counterparts positioned at galaxy centers (Fig.~\ref{fig_sample}).

One object from the final sample, SDSS J171409.04\-+584906.2, has archival \emph{Hubble Space Telescope} images. We observed 4 new confirmed IMBH host galaxies and 6 additional candidates with the Magellan Baade telescope using the \emph{FourStar} near-infrared imaging camera in the $K_s$ photometric band ($\lambda=2.2 \mu$m). 
We performed a light profile decomposition of IMBH host galaxy images, computed the luminosities of the spheroidal components and converted them into stellar masses using published ages and metallicities from RCSED.

\subsection{Optical and NIR observations}

We carried out follow-up imaging and spectroscopic observations of several IMBH candidates and their host galaxies in the optical and near-infrared domains using the 6.5-m Magellan Baade telescope, Las Campanas Observatory, Chile.

\begin{deluxetable*}{llccccc}
\tablecolumns{7}
\tablecaption{The log of follow-up observations of confirmed IMBHs and their host galaxies. \label{obslog}}

\tablehead{
	\colhead{Object} &
    \colhead{Instrument} &
    \colhead{Date} &
    \colhead{Exp. time} &
    \colhead{Seeing} \\
 	\colhead{} &
    \colhead{} &
    \colhead{} &
    \colhead{(s)} &
    \colhead{(arcsec)}
}

\startdata
J122732.18+075747.7 & MagE & 10/07/2017  & 3600 & 1.2\\
\hline
\multirow{3}{*}{J110731.23+134712.8} & MagE & 06/07/2017 & 2400& 1.3\\
 & FourStar & 10/07/2017 & 466 & 0.77\\
 & {\it Chandra} & 17/07/2017 & 9960 & n/a\\
\hline
\multirow{2}{*}{J134244.41+053056.1} & MagE & 30/05/2017 & 4800 & 1.5 \\
& FourStar & 09/07/2017  & 384 & 0.53\\
\hline
\multirow{2}{*}{J022849.51$-$090153.8} & MagE & 01/01/2018 & 3600 &0.9\\
& FourStar & 01/01/2018  & 384 & 0.7\\
\enddata

\end{deluxetable*}

Our primary goal was to obtain quasi-simultaneous optical spectroscopy of the IMBH galaxies selected for follow-up X-ray observations using {\it Chandra} and {\it XMM-Newton} within a period of 2--6 weeks between observations. Our secondary goal was to obtain the second spectroscopic epoch for several prominent X-ray confirmed IMBHs and get an independent black hole mass estimates. Finally, we aimed to take advantage of superior seeing conditions at the Magellan telescope to obtain near-infrared images of several IMBH host galaxies with the spatial resolution 0.5--0.7~arcsec crucial for the analysis of structural properties, that is 2--3 times better than the resolution of SDSS images. The complete log of our follow-up observations for confirmed IMBHs is provided in Table~\ref{obslog}.

For our spectroscopic observations, we used the Magellan Echellette spectrograph \citep{2008SPIE.7014E..54M} with the 10-arcsec long 0.7~arcsec wide slit that provides a cross-dispersion spectroscopy with the spectral resolving power $\lambda/\Delta \lambda = 6500$ or $\sigma_{\mathrm{inst}}=20$~km~s$^{-1}$ in 14 spectral orders covering the wavelength range $0.3<\lambda<1.0 \mu$m. Each object was integrated for 40~min to 1~h~20~min in individual 20~min long exposures either along the major or minor axis of its host galaxy. Objects which were small enough to fit in the slit ($<$5~arcsec) were observed along the minor axis and the sky model was constructed from the ``empty'' part of the slit. Larger galaxies which did not fit in a 10~arcsec slit were observed along the major axis; then we used offset sky observations of 5~min re-normalized in flux to match science observations.

We reduced the data using a dedicated MagE data reduction pipeline, which we developed. The pipeline builds a wavelength solution with uncertainties as small as 2~km~s$^{-1}$; merges echelle orders and creates a flux calibrated sky subtracted merged 2D spectrum, which is then fed to the {\sc NBursts} spectral fitting procedure to subtract the stellar continuum and then to the emission line analysis procedure. 

The pipeline uses standard stars observed shortly before or after a science source to perform flux calibration and telluric correction. However, in order to perform an extra check and eliminate possible systematic flux calibration errors, we compare our reduced spectra to SDSS and use stellar continuum to perform independent flux calibration. We first extract a spectrum in a 3$\times$0.7~arcsec box and use a published light profile of a each galaxy \citep{simard11} in order to calculate the expected flux difference between the circular 3~arcsec SDSS fiber aperture and the box extraction. Then, we calibrate a reduced 2D spectrum using that flux ratio, and perform an optimal extraction of a nuclear point source using the value of the image quality reported by the guider in order to estimate an extraction profile, because the BLR in an AGN is supposed to stay unresolved. At the end, we apply a flux correction computed for a point source observed through a 0.7~arcsec wide slit to the extracted spectrum. This approach yields a flux calibrated spectrum of the galaxy nucleus that can be directly compared to SDSS.

For our imaging observations we used the FourStar camera \citep{2013PASP..125..654P} that covers a field of view of 11$\times$11~arcmin with a mosaic containing 4 Hawaii2-RG detectors. We observed each IMBH host galaxy from a subsample selected for imaging in the $K_s$ band with the total on-source integration of 8~min. We used the Poisson random dithering pattern with a box size of 52~arcsec in order to provide enough background sampling for flat fielding and background subtraction. We reduced FourStar images using the {\sc fsred} data reduction pipeline that performs pre-processing of raw NIR images obtained in the \emph{fowler2} mode, that is 2 read-outs in the beginning and 2 at the end of each exposure; flat fielding; background subtraction; and flux calibration using 2MASS \citep{skrutskie06} sources inside the field of view. The final result of the pipeline is a sky subtracted flux calibrated image and its flux uncertainties.

For one galaxy, SDSS~J1714+5849, we used archival Hubble Space Telescope images in the F814W photometric band downloaded from the Hubble Legacy Archive (\url{http://hla.stsci.edu/}; dataset JA2S0M010). For SDSS~J1227+0757, which we did not observe with FourStar, we used Pan-STARRS archival images \citep{2016arXiv161205560C} with sub-arcsec seeing quality.

In order to check the position of all our candidates on the $M_{\mathrm{BH}}-M_{\mathrm{bulge}}$ scaling relation, we analyzed imaging data for the candidate IMBH host galaxies. For all of them but one we have used a two-dimensional decomposition using the {\sc galfit v.3} software \citep{Peng10}. For one galaxy, SDSS~J1714+5849, which harbors a strong stellar bar, we used instead a one-dimensional decomposition of a light profile extracted using the {\sc ellipse} task in {\sc noao iraf}.
For SDSS~J1342+0530, SDSS~J1107+1347, and SDSS~J0228$-$0901 the follow-up imaging data taken on FourStar has been used. We used {\sc psfextractor} \citep{bertin11} to extract the point spread function convolution kernel for every image.
Usually we found the best-fitting solution with the simple photometric model ``bulge+disk''. In case of compact bulges being limited by the atmospheric seeing quality we had to model a bulge using a point source.
Then, apparent magnitudes of bulges were translated into luminosities using the WMAP9 cosmology \citep{hinshaw13} and later converted into stellar masses using stellar population properties provided in the RCSED \citep{RCSED}.

\subsection{X-ray data and observations analysis}

We performed X-ray observations of 2 objects with {\it Chandra} (observations 20114 and 20115), and 2 objects with {\it XMM-Newton} (observations 0795711301 and 0795711401) using director's discretionary time quasi-simultaneously with optical observations.

Both {\it Chandra} observations were carried out with the Advanced CCD Imaging Spectrometer (ACIS) detector in the faint data mode with 10,000~s long exposures.
Target galaxies were always placed on-axis of the back-illuminated S3 chip of ACIS-S.
The data were reduced and analyzed with {\sc ciao 4.9} package following standard recipes.

In {\it Chandra} observation 20114 a single bright point-like X-ray source was detected at the position of the optical center of SDSS~J1107+1347 galaxy.
We performed its aperture photometry using {\sc srcflux} task and detected 518 net counts in 0.5--7~keV band which corresponds to the observed flux $(4.5 \pm 0.4) \times 10^{-13}$~\ergscm.

No source was detected at the position of SDSS J135750.71+223100.8 in the {\it Chandra} observation 20115.
We estimate a $3\sigma$ detection limit of this observation as $8.0 \times 10^{-15}$~\ergscm.

Two {\it XMM-Newton} observations 10,000~s long each were performed with EPIC detector in FullFrame mode with Thin filter.
The data were reduced and analyzed with the common {\it XMM-Newton} analysis threads with the {\sc sas} 16.1.0 software package running in a virtual machine.

We were not able to detect any source at a position of SDSS~J161251.77+110621.6 in the {\it XMM-Newton} observation 0795711301 up-to the limiting flux level of $5.0 \times 10^{-15}$~\ergscm.

However, we marginally detected a faint source coincident within positional errors with the nucleus of SDSS~J1440+1155 galaxy in the {\it XMM-Newton} observation 0795711401.
We estimate its flux in the standard {\it XMM-Newton} 0.2--10~keV band as $(5 \pm 2) \times 10^{-15}$~\ergscm.

For other sources in this study (including previously known objects) we used X-ray data from {\it XMM-Newton}'s 3XMM-DR7 catalog \citep{rosen16} accessible through the catalog website \citep[\url{http://xmm-catalog.irap.omp.eu}]{zolotukhin17}, {\it Chandra} Source Catalog Release 1.1 \citep{evans10} and {\it Swift} 1SXPS catalog \citep{evans14}.
We performed a cross-match with point non-spurious subset of sources in those catalogs using their 3$\sigma$ X-ray positional uncertainties.

\section{Results and Discussion}

\subsection{Detected IMBH candidates and their properties}

\begin{deluxetable*}{lcccccccccc}
\tiny
\tablecolumns{10}
\tablecaption{IMBHs identified as AGN and some of their properties.}

\tablehead{
	\colhead{Object} &
    \colhead{$M_{\rm BH}$} &
    \colhead{Lit.$M_{\rm BH}$} &
    \colhead{$\sigma_{\rm BLR}$} &
    \colhead{$L_{\rm BLR} $ H$\alpha$} &
    \colhead{z} &
    \colhead{$M_{\rm abs}^{\rm sph}$} &
    \colhead{$M_{\rm sph}^{*}$} &
    \colhead{$L_{\rm X}$} \\
 	\colhead{} &
    \colhead{($10^3 M_{\odot}$)} &
    \colhead{($10^3 M_{\odot}$)} &
    \colhead{(km s$^{-1}$)} &
    \colhead{($10^{39}$ erg s$^{-1}$)} &
    \colhead{} &
    \colhead{(mag)} &
    \colhead{($ 10^9 M_{\odot}$)} &
    \colhead{($10^{40}$ erg s$^{-1}$)}
}

\startdata
\multicolumn{9}{c}{This work}\\
\hline
\multirow{2}{*}{\object[SDSS J122732.18+075747.7]{J122732.18+075747.7}} & $43\pm10$\tablenotemark{1}  & \multirow{2}{*}{  } & $214\pm20$ & $1.5\pm0.4$ & \multirow{2}{*}{ 0.033 } & \multirow{2}{*}{ $-$17.8 (r) } & \multirow{2}{*}{ 0.9 } & \multirow{2}{*}{ 0.55 ({\it XMM}) } \\
 & $36\pm7$\tablenotemark{2} & & $200\pm10$ & $1.4\pm0.4$ & & & & \\
\hline
\multirow{2}{*}{\object[SDSS J134244.41+053056.1]{J134244.41+053056.1}} & $65\pm7$\tablenotemark{1} & \multirow{2}{*}{  } & $216\pm10$ & $3.5\pm0.4$ & \multirow{2}{*}{ 0.037 } & \multirow{2}{*}{ $-$20.7 (r) } & \multirow{2}{*}{ 3.5 } & \multirow{2}{*}{ 13.5 ({\it Swift}) } \\
 & $96\pm13$\tablenotemark{2} & & $286\pm13$ & $2.4\pm0.5$ & & & & \\
\hline
\object[SDSS J171409.04+584906.2]{J171409.04+584906.2} & $115\pm24$\tablenotemark{1} &  & $373\pm31$ & $1.1\pm0.3$ & 0.030 & $-$17.4 (F814W) & 0.7 & 2.5 ({\it XMM}) \\
\hline
\object[SDSS J111552.01-000436.1]{J111552.01$-$000436.1} & $115\pm38$\tablenotemark{1} &  & $315\pm41$ & $2.3\pm0.9$ & 0.039 & $-$16.8 (r) & 0.4 & 4.9 ({\it XMM}) \\
\hline
\multirow{2}{*}{\object[SDSS J110731.23+134712.8]{J110731.23+134712.8}} & $122\pm18$\tablenotemark{1} & \multirow{2}{*}{  } & $269\pm17$ & $5.1\pm0.8$ & \multirow{2}{*}{ 0.045 } & \multirow{2}{*}{ $-$18.0 (K) } & \multirow{2}{*}{ 0.3 } & \multirow{2}{*}{ 190 ({\it Chandra})$^\star$ } \\
 & $71\pm10$\tablenotemark{2} & & $244\pm10$ & $2.5\pm0.6$ & & & & \\
\hline
\multicolumn{9}{c}{Previously known}\\
\hline
\object[SDSS J152304.97+114553.6]{J152304.97+114553.6}\tablenotemark{a} & $70\pm20$\tablenotemark{1} & 50 & $350\pm30$ & $0.5\pm0.2$ & 0.024 &  & 0.7 & 0.4 ({\it Chandra})\tablenotemark{a} \\
\hline
\object[SDSS J153425.58+040806.7]{J153425.58+040806.7}\tablenotemark{b} & $111\pm7$\tablenotemark{1} & 130 & $246\pm6$ & $6.2\pm0.3$ & 0.039 &  & 1.3 & 85 ({\it Chandra})\tablenotemark{d} \\
\hline
\object[SDSS J160531.84+174826.1]{J160531.84+174826.1}\tablenotemark{b} & $116\pm11$\tablenotemark{1} & 160 & $316\pm12$ & $2.3\pm0.2$ & 0.032 &  & 1.7 & 12.7 ({\it XMM}) \\
\hline
\object[SDSS J112333.56+671109.9]{J112333.56+671109.9}\tablenotemark{c} & $157\pm36$\tablenotemark{1} & 200 & $341\pm34$ & $3.1\pm0.6$ & 0.055 &  & 2.4 & 53.5 ({\it XMM}) \\
\hline
\multirow{2}{*}{\object[SDSS J022849.51$-$090153.8]{J022849.51-090153.8\tablenotemark{c}}} & $202\pm13$\tablenotemark{1} & \multirow{2}{*}{ 316 } & $250\pm7$ & $21\pm1$ & \multirow{2}{*}{ 0.072 } & \multirow{2}{*}{  } & \multirow{2}{*}{ 0.7 } & \multirow{2}{*}{ 275 ({\it Chandra})\tablenotemark{d} } \\
 & $367\pm27$\tablenotemark{2} & & $340\pm9$ & $19\pm2$ & & & & \\
\hline
\hline
\enddata

\tablenotetext{1}{Spectrum from SDSS}
\tablenotetext{2}{Spectrum from Magellan/MagE}
\tablenotetext{a}{\citet{baldassare15}}
\tablenotetext{b}{\citet{reines13}}
\tablenotetext{c}{\citet{2007ApJ...670...92G}}
\tablenotetext{d}{\citet{dong12}}
\tablecomments{The upper part of the table lists objects found in this work, while 5 objects in the bottom part were known previously (see references near object names) but had their properties re-measured using our data analysis approach. The X-ray luminosity is computed in this work from the flux reported in a corresponding X-ray catalog unless a reference is given. An asterisk marks dedicated {\it Chandra} X-ray observations from this study. The columns are: SDSS IAU name, black hole mass (derived in this study from SDSS data and from Magellan/MagE data where available), black hole mass from the literature (where applicable; see reference near object name), redshift, H$\alpha$ BLR velocity dispersion (from SDSS and from MagE data where available), H$\alpha$ BLR luminosity (from SDSS and from MagE data where available), absolute magnitude and mass of spheroidal component (bulge) of a host galaxy, X-ray luminosity.\label{tableimbh}}
\end{deluxetable*}

\begin{deluxetable*}{cccccc}
\tablecolumns{6}
\tablecaption{A list of 305 candidate IMBHs identified as active galactic nuclei based on the SDSS archival data.}

\tablehead{
	\colhead{Object} &
    \colhead{z} &
    \colhead{$M_{\rm BH}$} &
    \colhead{$\sigma_{\rm BLR}$} &
    \colhead{$L_{\rm BLR} $ H$\alpha$} &
    \colhead{$M_{\rm abs}^{\rm sph}$} \\
 	\colhead{} &
    \colhead{} &
    \colhead{($10^3 M_{\odot}$)} &
    \colhead{(km s$^{-1}$)} &
    \colhead{($10^{39}$ erg s$^{-1}$)} &
    \colhead{($10^9 M_{\odot}$)}
}

\startdata
\object[SDSS J111835.82+002511.2]{J111835.82+002511.2} & 0.025 & 138 $\pm$ 20 & 230 $\pm$ 13 & 13.24 $\pm$ 2.07 & 1.87 $\pm$ 0.04 \\
\object[SDSS J112209.97+010114.8]{J112209.97+010114.8} & 0.075 & 99 $\pm$ 19 & 216 $\pm$ 14 & 86.3 $\pm$ 2.51 & 1.94 $\pm$ 0.3 \\
\object[SDSS J141215.60-003759.0]{J141215.60$-$003759.0} & 0.025 & 62 $\pm$ 17 & 269 $\pm$ 30 & 1.24 $\pm$ 0.39 & 0.46 $\pm$ 0.05 \\
\object[SDSS J094733.06+001302.9]{J094733.06+001302.9} & 0.063 & 181 $\pm$ 39 & 322 $\pm$ 30 & 5.47 $\pm$ 1.15& 10.65 $\pm$ 2.50 \\
\object[SDSS J003826.70+000536.8]{J003826.70+000536.8} & 0.071 & 100 $\pm$ 22 & 257 $\pm$ 25 & 4.06 $\pm$ 0.87 & n/a \\
\enddata

\tablecomments{This table is published in its entirety in the machine-readable format.
A random subset is shown here for guidance regarding its form and content.
\label{table_all_objs}}
\end{deluxetable*}

\begin{figure*}
\centering
\includegraphics[width=0.8\textwidth]{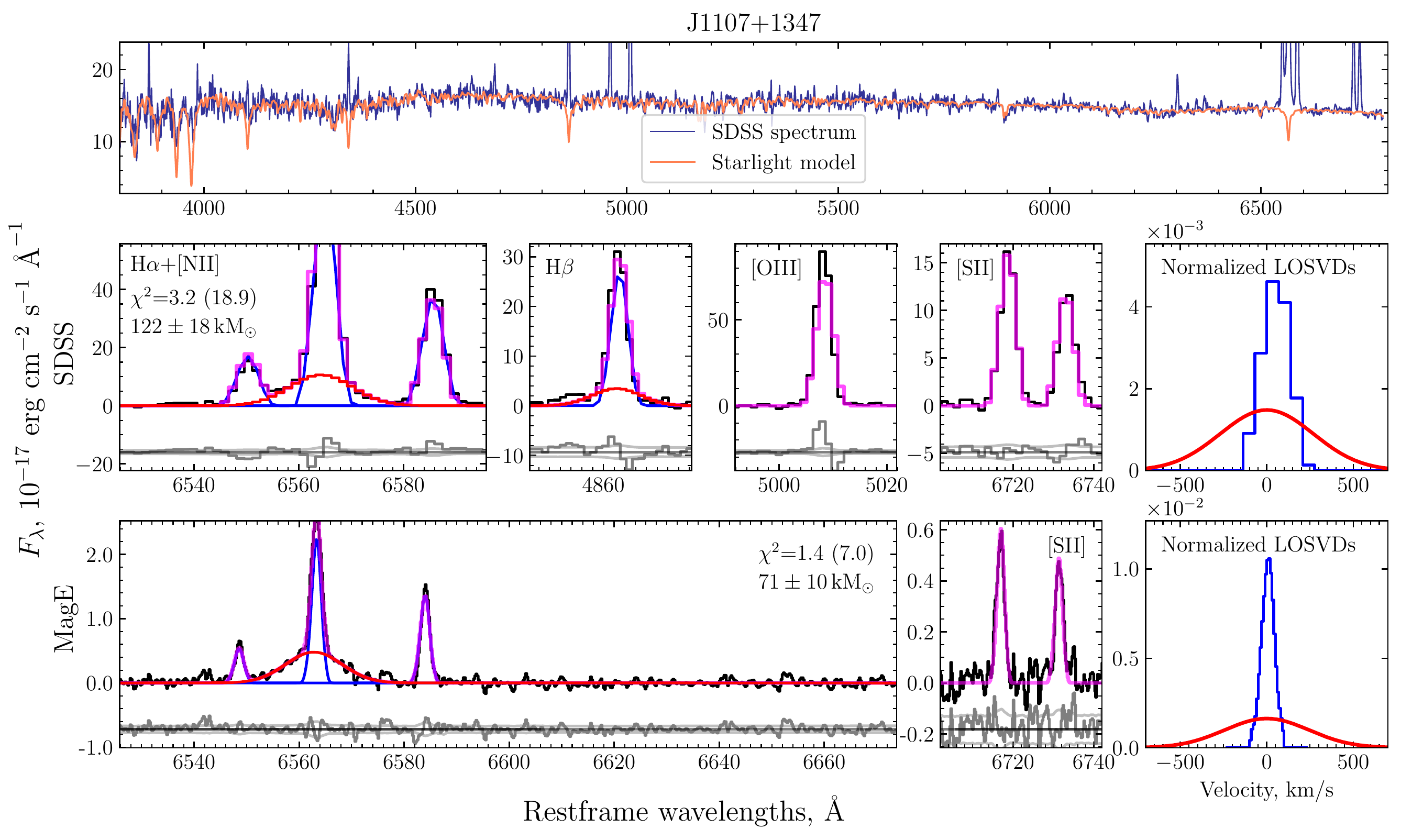}\\
\caption{Spectral decomposition of MagE and SDSS data for IMBHs detected in this work.
Top row panel: The observed optical SDSS spectrum is shown in blue, the orange line is the best-fitting stellar population template without emission lines.
Middle row: A close-up view of several emission lines in the SDSS spectrum.
The emission line profile (observed data are shown in black) is constructed first by subtracting the best-fitting stellar population template, then in allowed lines it is decomposed into narrow line (gray) and broad line (red) components.
The total emission line model is shown in magenta and its residuals are displayed in gray shifted downward for clarity.
A non-parametric narrow line model (rightmost panel, blue histogram) is computed simultaneously for all forbidden emission lines in the spectrum that reduces fitting residuals and allows us to detect even a very faint broad line component (rightmost panel, red line), a proxy for the BH mass.
Bottom row: Same as the middle row but for MagE data.
\label{fig_decomposition_our_mage_1}}
\end{figure*}

\setcounter{figure}{2}
\begin{figure*}
\centering
\includegraphics[width=0.8\textwidth]{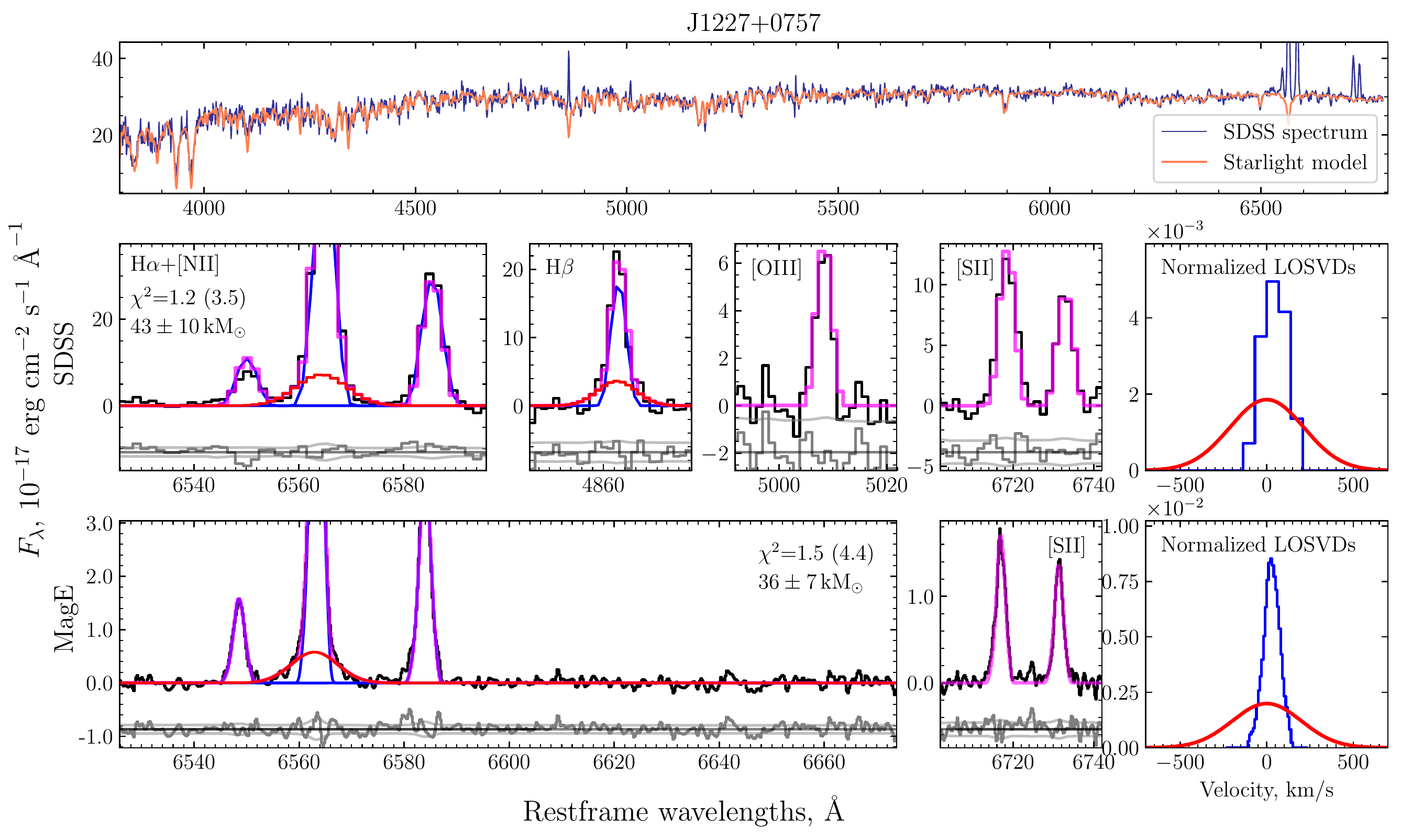}\\
\includegraphics[width=0.8\textwidth]{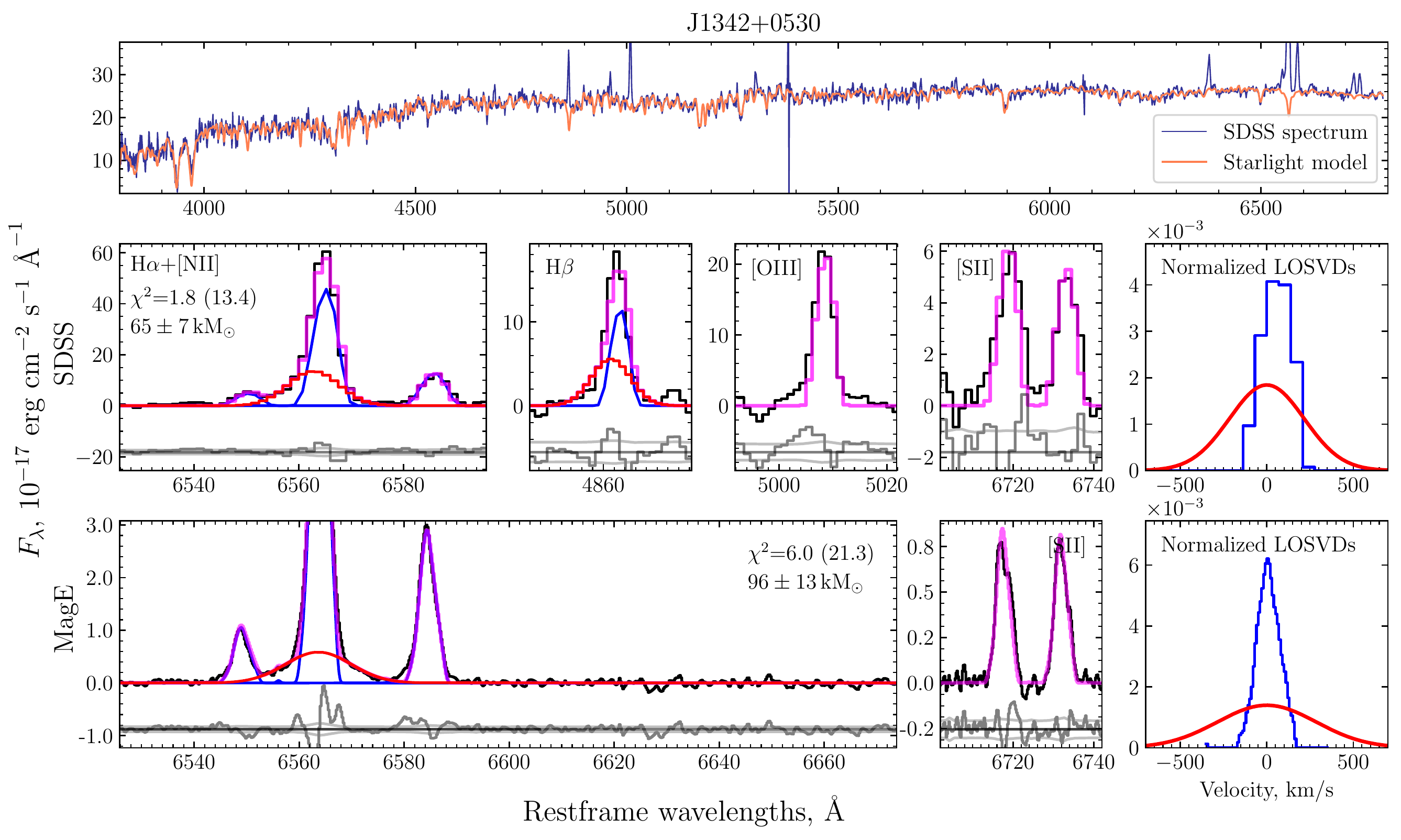}
\caption{contd.
\label{fig_decomposition_our_mage_2}}
\end{figure*}

\begin{figure*}
\centering
\includegraphics[width=0.8\textwidth]{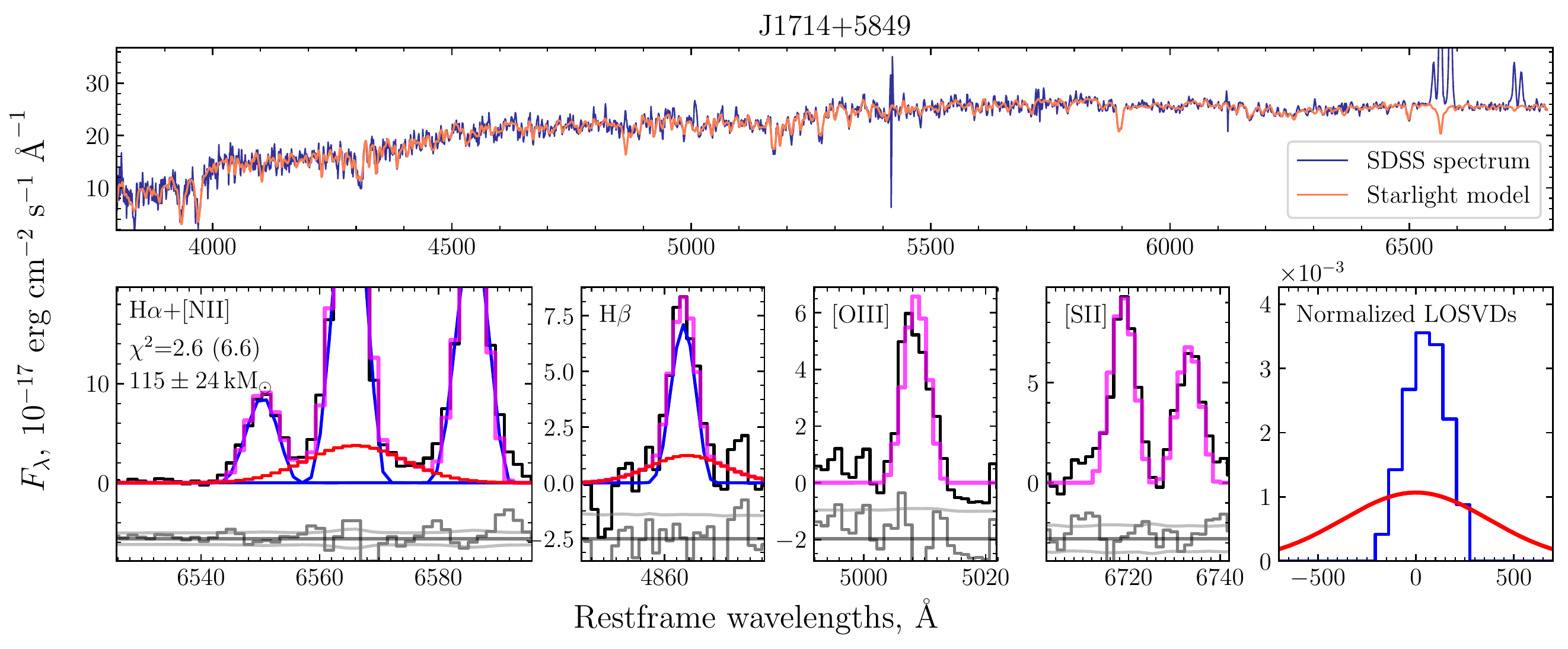}\\
\includegraphics[width=0.8\textwidth]{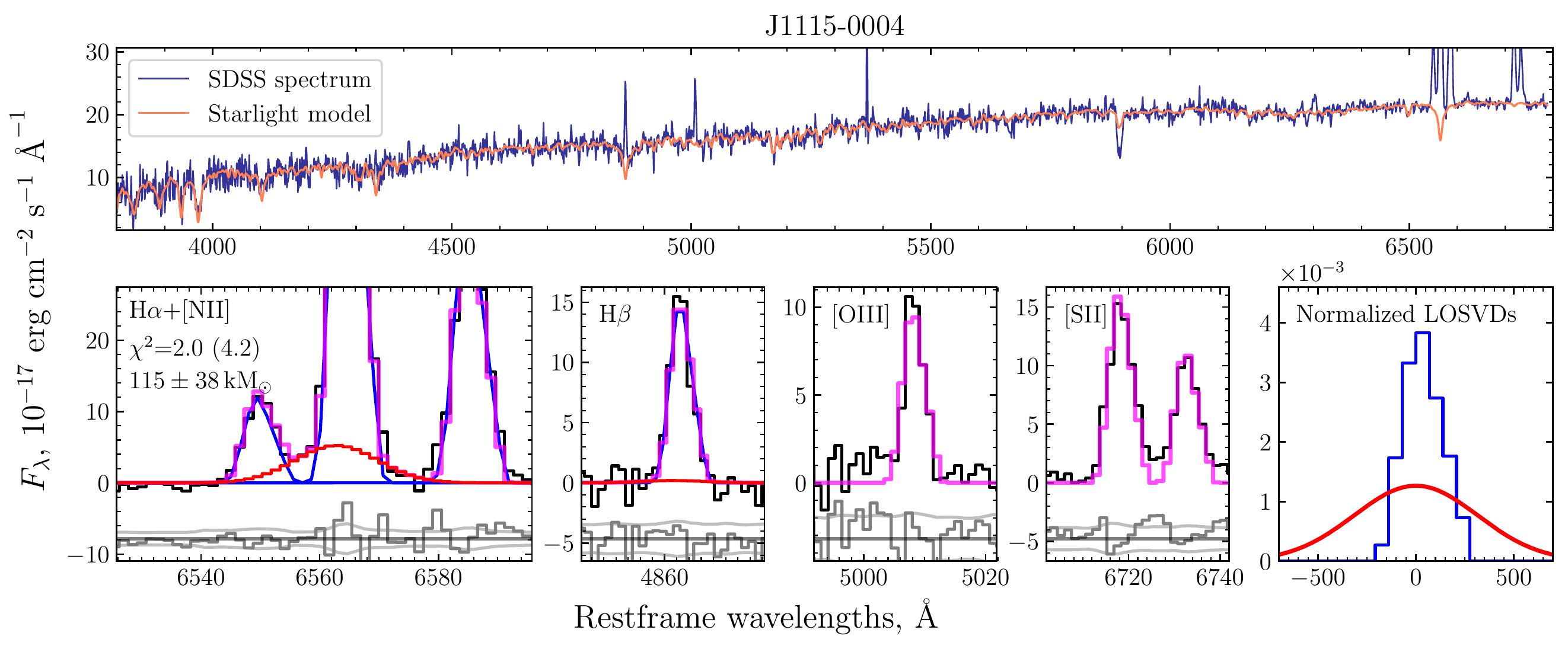}\\
\caption{Spectral decomposition of SDSS data for IMBHs detected in this work.
Panels are the same as top and middle rows in Figure~\ref{fig_decomposition_our_mage_1}.
\label{fig_decomposition_our}}
\end{figure*}

\begin{figure*}
\centering
\includegraphics[width=0.8\textwidth]{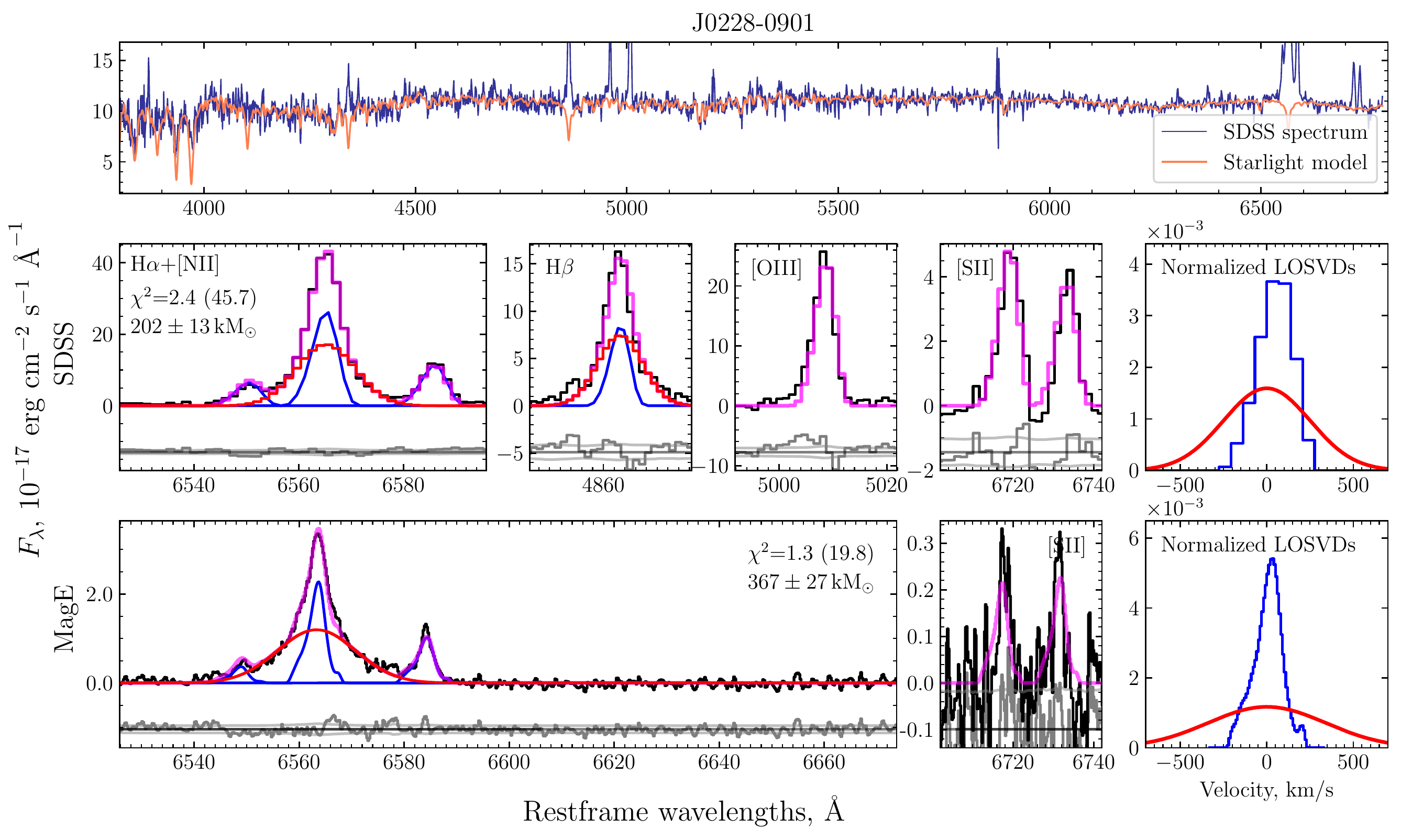}
\caption{Spectral decomposition of MagE and SDSS data for a previously known IMBH re-measured in this work.
Panels are the same as in Figure~\ref{fig_decomposition_our_mage_1}.
\label{fig_decomposition_lit_mage}}
\end{figure*}

\begin{figure*}
\centering
\includegraphics[width=0.8\textwidth]{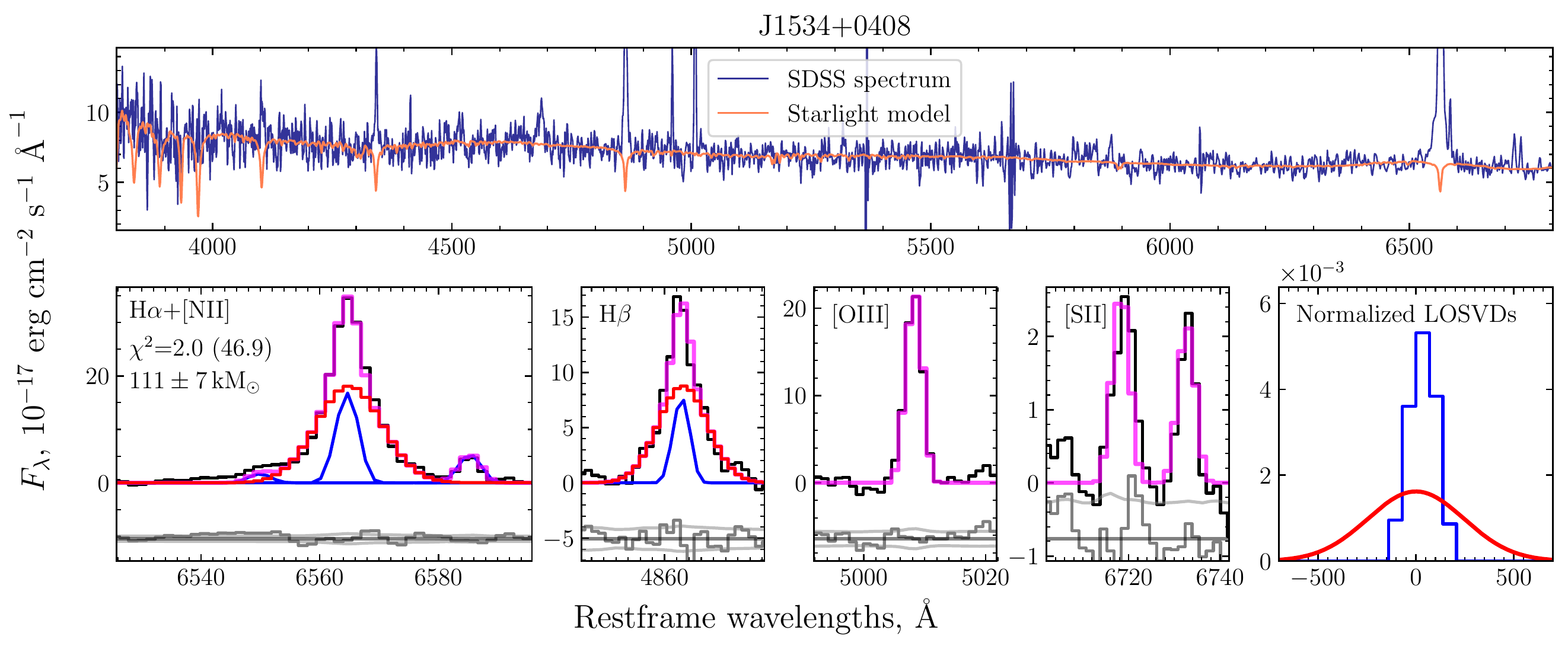}\\
\includegraphics[width=0.8\textwidth]{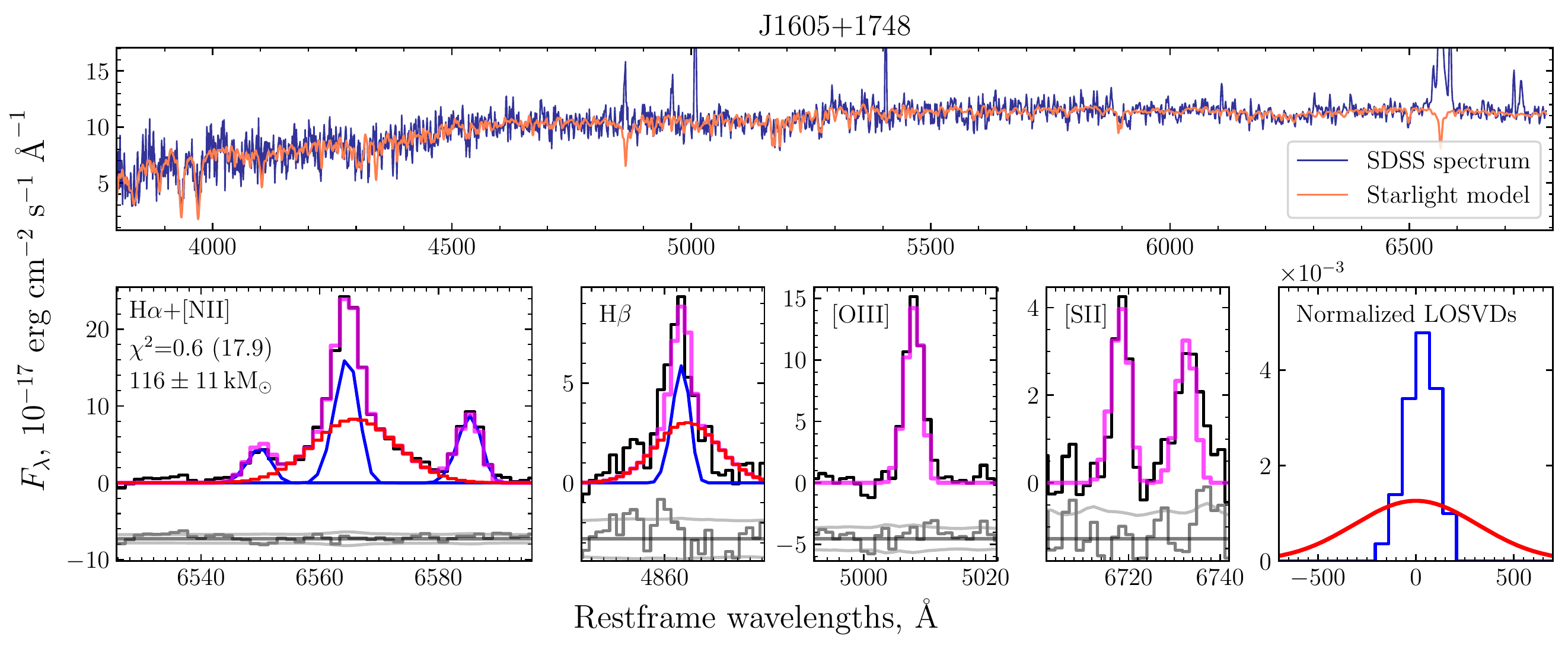}\\
\includegraphics[width=0.8\textwidth]{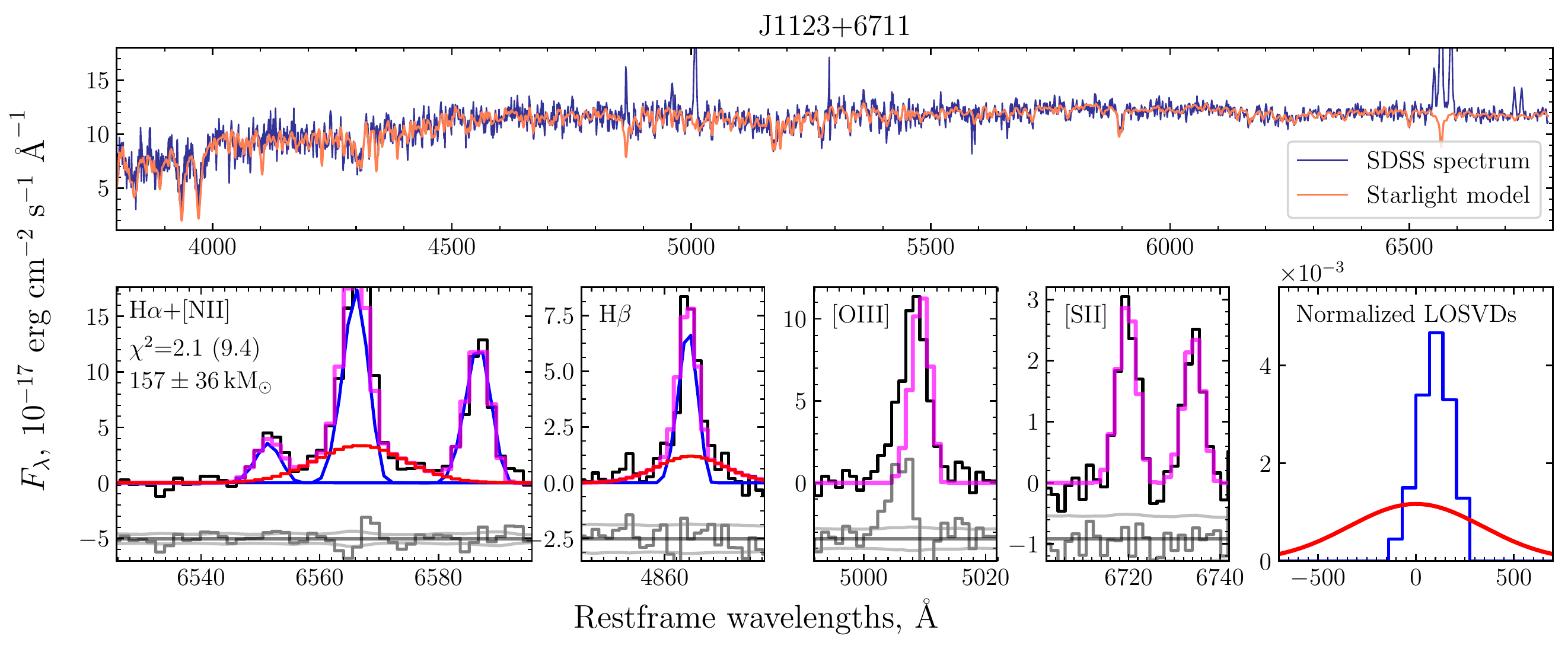}\\
\caption{Spectral decomposition of SDSS data for previously known IMBHs re-measured in this work.
Panels are the same as in Figure~\ref{fig_decomposition_our}.
\label{fig_decomposition_lit}}
\end{figure*}

\setcounter{figure}{5}
\begin{figure*}
\centering
\includegraphics[width=0.8\textwidth]{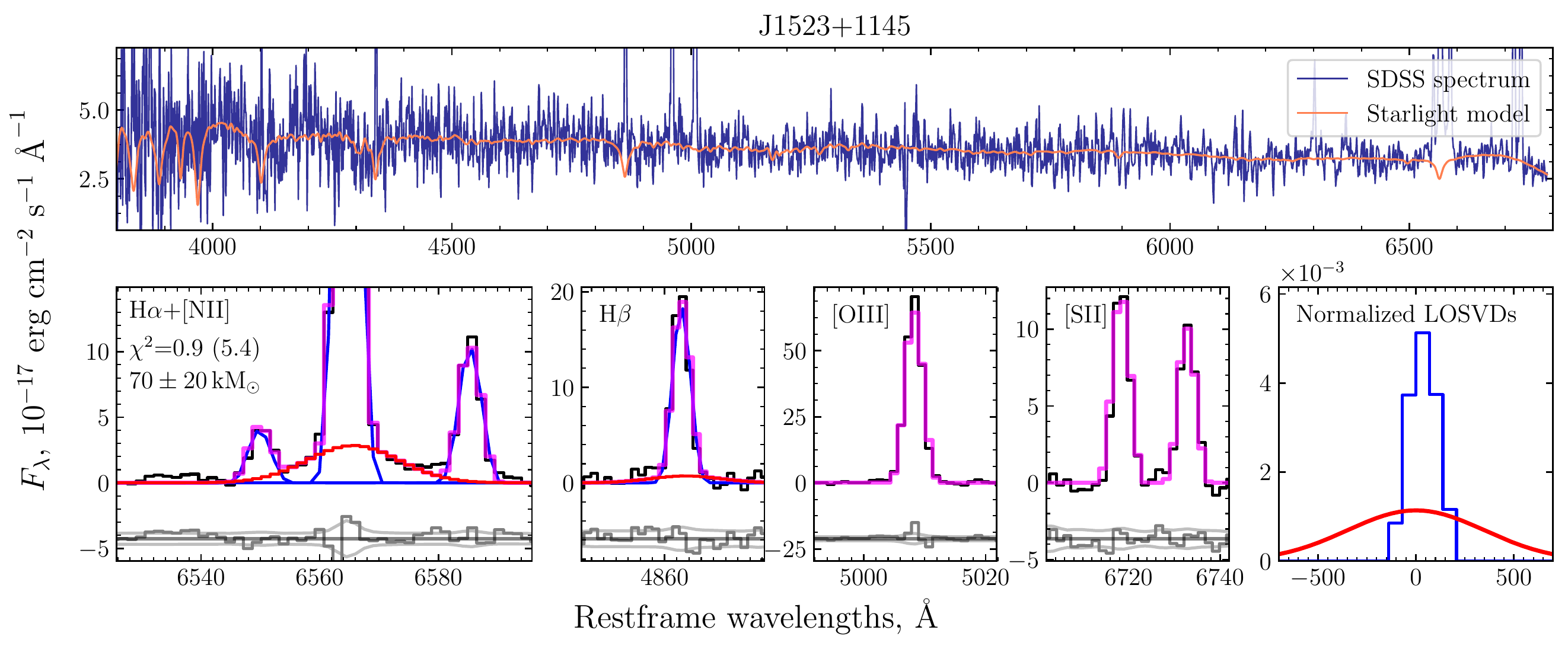}
\caption{contd.
\label{fig_decomposition_our_mage_3}}
\end{figure*}

Using data mining in wide-field sky surveys and applying dedicated analysis to archival and follow-up optical spectra, we identified a sample of 305 IMBH candidates having masses $3\times10^4<M_{\mathrm{BH}}<2\times10^5 M_{\odot}$, which reside in galaxy centers and are accreting gas that creates characteristic signatures of a type-I AGN. We confirmed the AGN nature of ten sources (including five previously known \citep{dong07,reines13,baldassare15}) by detecting the X-ray emission from their accretion discs, thus defining the first \emph{bona fide} sample of IMBHs in galactic nuclei. The very existence of nuclear IMBHs supports the stellar mass seed scenario of the massive black hole formation.

In Table~\ref{tableimbh} we present main properties of 10 IMBHs confirmed as AGN by the X-ray identification and their host galaxies. Every object is identified by the IAU designation, which includes its {\it J2000} coordinates. For every source we present a central BH virial mass estimate, flux and width of the broad H$\alpha$ component, redshift, X-ray flux, and an estimated stellar mass of a spheroidal component. For host galaxies of 5 newly detected sources presented in the top part of the table, we also provide estimates of the absolute magnitude of the bulge or spheroid obtained from the photometric decomposition of their direct images. For the confirmed sources from the literature (bottom part of the table) we also provide published BH mass estimates. In Fig.~\ref{fig_decomposition_our_mage_1}--\ref{fig_decomposition_our_mage_3} we present SDSS and MagE (when available) spectra and line profile decomposition results for these 10 objects.

Here we briefly describe properties of \emph{bona fide} IMBHs detected in X-ray for the first time:
\begin{itemize}
\item \object[SDSS J122732.18+075747.7]{J122732.18+075747.7}: The least massive IMBH ($M_{\rm BH}=3.6\times10^4M_\odot$) detected by our workflow hosted in a barred spiral galaxy with a star-forming ring; the X-ray counterpart is very faint. The BPT diagnostics places a galaxy in the composite region because the AGN emission is heavily contaminated by star formation in the inner ring, that becomes less of a problem in MagE data where it is spatially resolved.
\item \object[SDSS J134244.41+053056.1]{J134244.41+053056.1}: A particularly interesting source, which we matched with the {\it Swift} source 1SXPS J134244.6+053052. \citet{dou16} classified it as a tidal disruption event (TDE) candidate based on the variability of highly ionized iron lines in its optical spectrum. The claim is that the TDE must have happened close to the SDSS spectrum epoch (2002-04-09) which is, however, in clear contradiction with the hypothesis that its X-ray emission with the luminosity $1.3 \times 10^{41}$~erg~s$^{-1}$ observed by {\it Swift} 7 years later on 2009-05-15 is connected to the TDE. Therefore we attribute the detected X-ray source to the AGN activity in SDSS~J1342+0530.
\item \object[SDSS J171409.04+584906.2]{J171409.04+584906.2}: Hosted in a barred spiral with a compact bulge well resolved in archival HST images, this IMBH is another example of a source falling into the composite region in the BPT diagram. Because of high declination we were unable to obtain the second epoch spectroscopy.
\item \object[SDSS J111552.01-000436.1]{J111552.01$-$000436.1}: Located in a nearly edge-on spiral galaxy with a compact bulge, this is another example of a weak AGN whose signature is contaminated by star formation in its host galaxy.
\item \object[SDSS J110731.23+134712.8]{J110731.23+134712.8}: This IMBH located in a low luminosity disk galaxy with a very compact bulge is the only one of five falling in the AGN region of the BPT diagram despite its contamination by star formation. This object has a very bright X-ray counterpart detected in our Chandra dataset, that corresponds to the X-ray luminosity alone over 10\%\ of the Eddington limit for a 70,000~$M_{\odot}$ black hole, which suggests that the bolometric luminosity should be close to the Eddington limit.
\end{itemize}

We also mention one object previously described in the literature, \object[SDSS J022849.51-090153.8]{J022849.51$-$090153.8} \citep{2007ApJ...670...92G}. Its black hole mass estimate from the follow-up spectroscopy with MagE, 3.7$\pm$0.3$\cdot 10^5 M_{\odot}$ puts it above the IMBH mass threshold adopted in this work, however, similarly to \object[SDSS J110731.23+134712.8]{J110731.23+134712.8} it also exhibits very bright X-ray emission, that corresponds to the bolometric luminosity close to the Eddington limit for its mass.

Table~\ref{table_all_objs} contains properties of all 305 sources selected as IMBH candidates ({\textit{the parent sample}}) regardless of the availability of X-ray data. The columns are similar to Table~\ref{tableimbh} with the exception of X-ray identification and literature data. We used bulge luminosities in the $r$ band reported in the photometric catalog by \citet{simard11} in the ``Sersic+disk'' decomposition table, which we converted into stellar masses using mass-to-light ratios of SDSS galaxies presented in \citet{saulder+16}.

\subsection{The IMBH mass detection limit and reliability of $M_{\rm BH}$ estimates}

\begin{figure*}
\centerline{
\includegraphics[clip,trim={0.0cm 0 2.25cm 0},height=0.199\hsize]{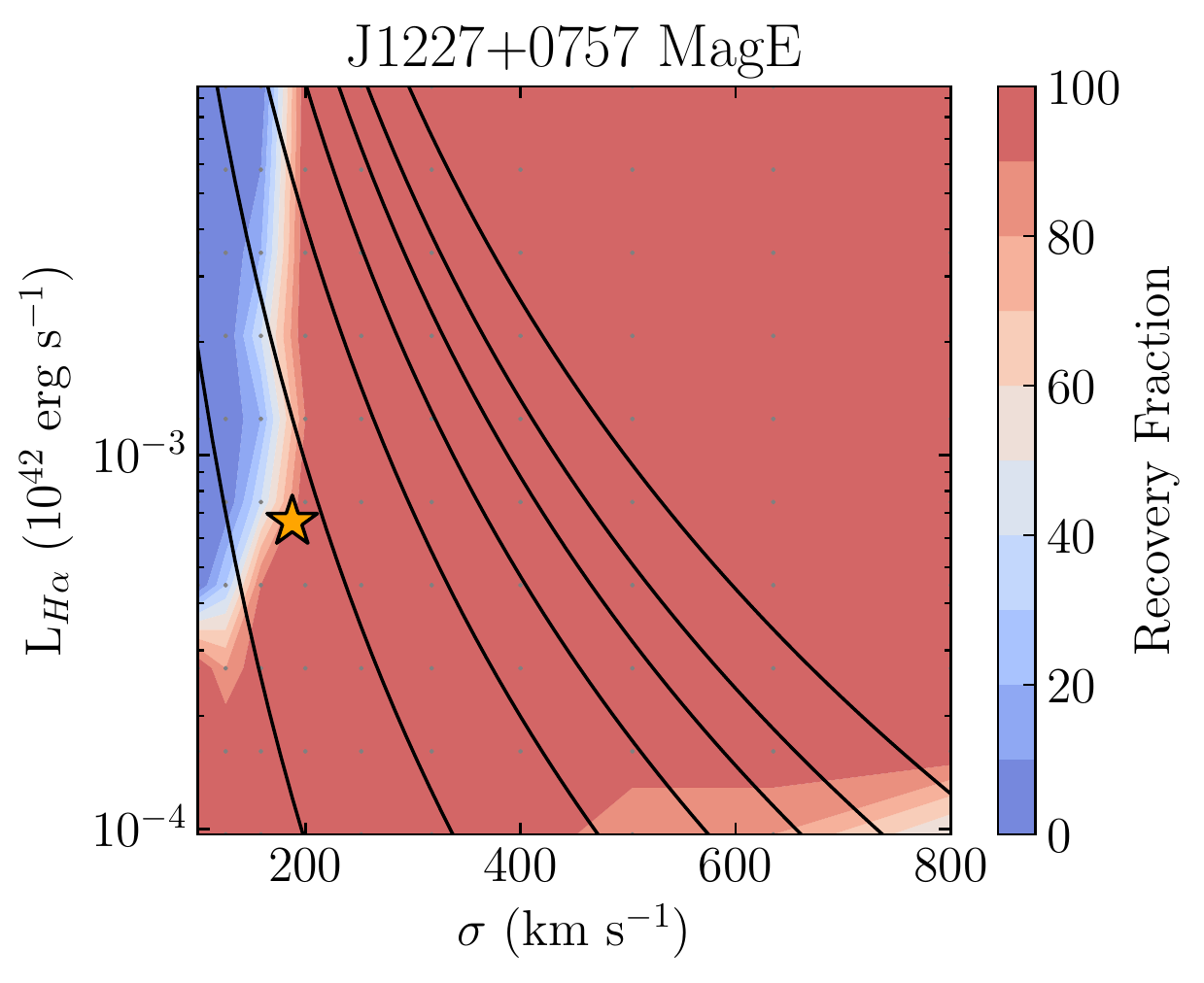} 
\includegraphics[clip,trim={1.04cm 0 2.25cm 0},height=0.199\hsize]{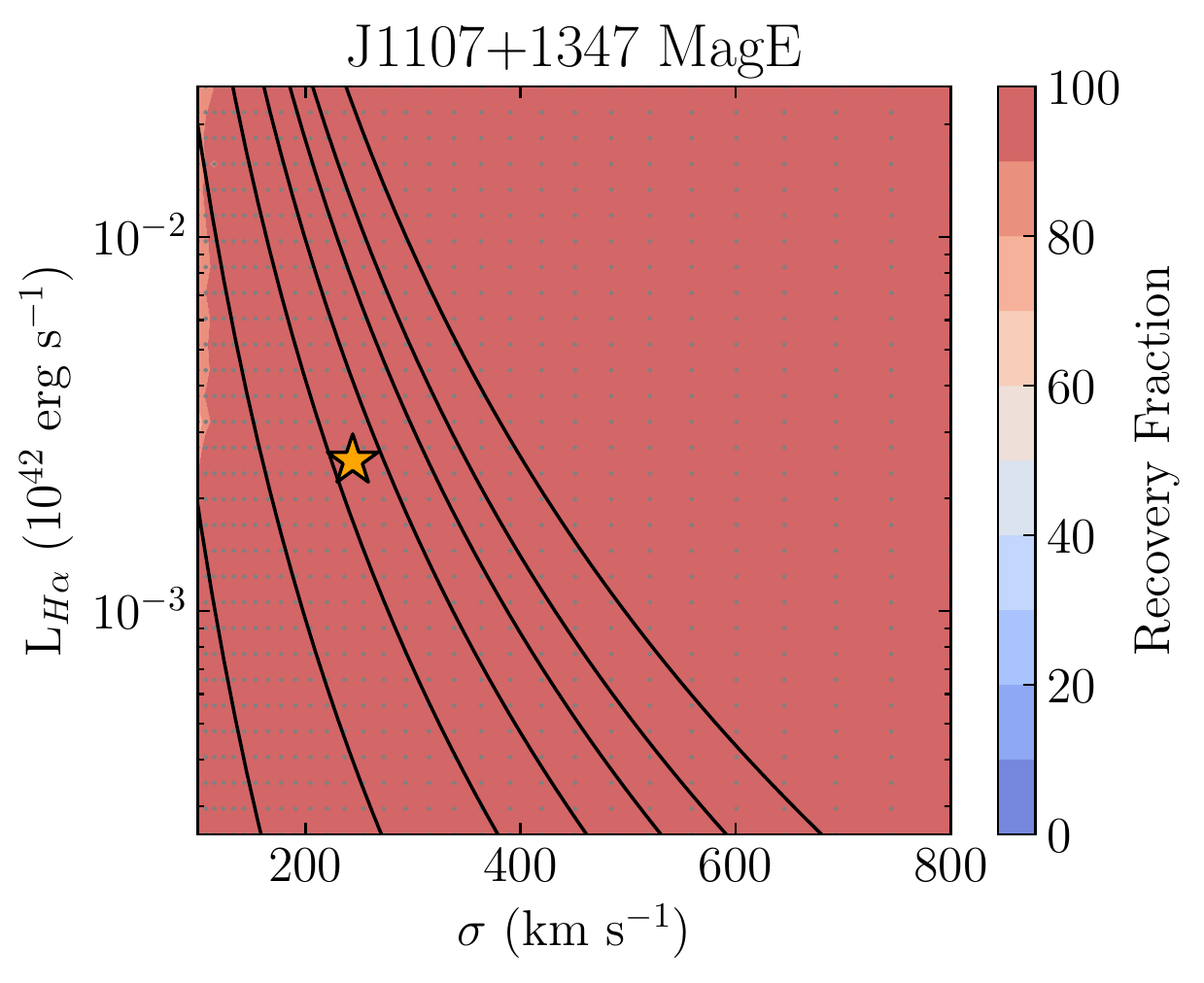} 
\includegraphics[clip,trim={1.04cm 0 2.25cm 0},height=0.199\hsize]{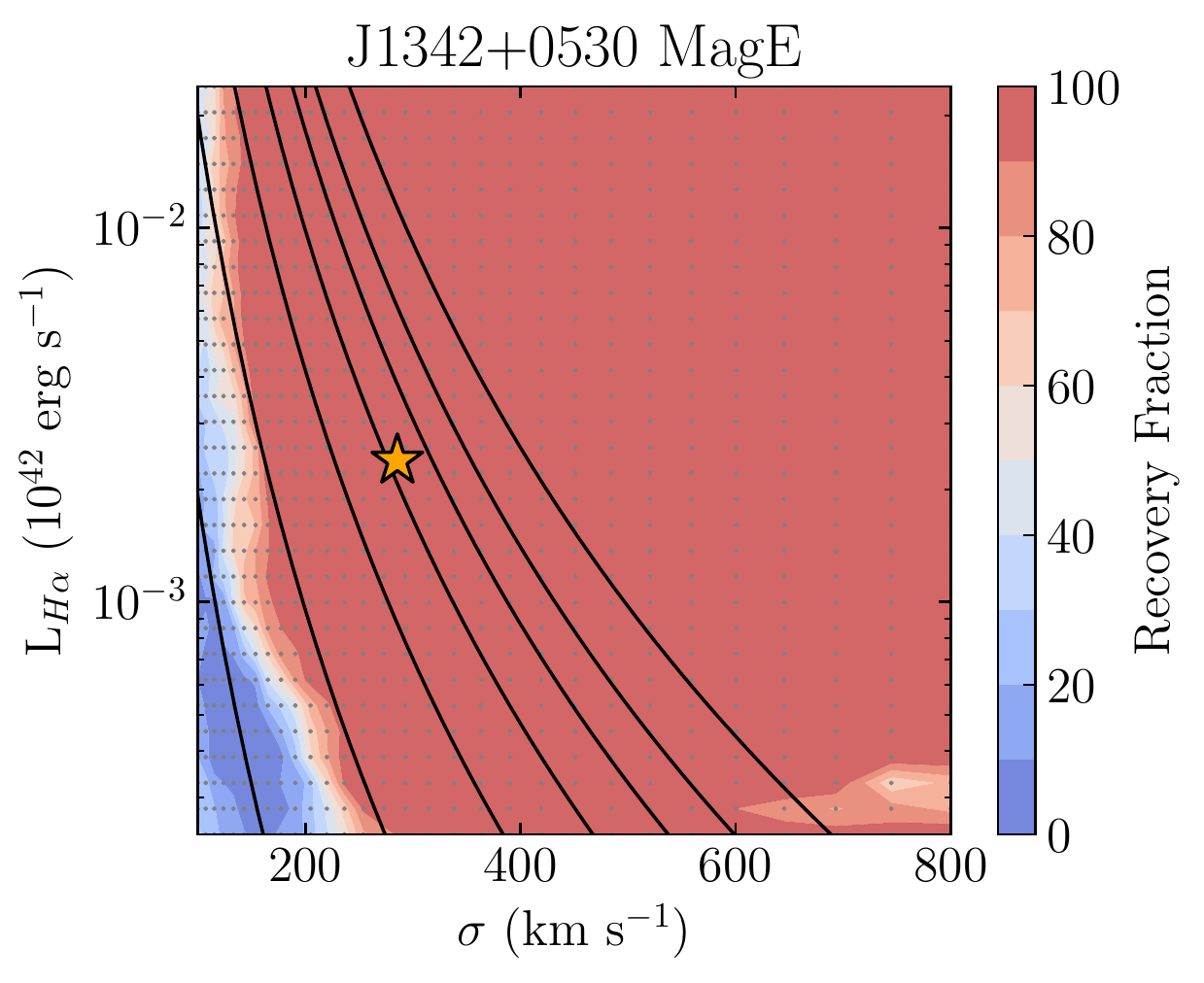} 
\includegraphics[clip,trim={1.04cm 0 2.25cm 0},height=0.199\hsize]{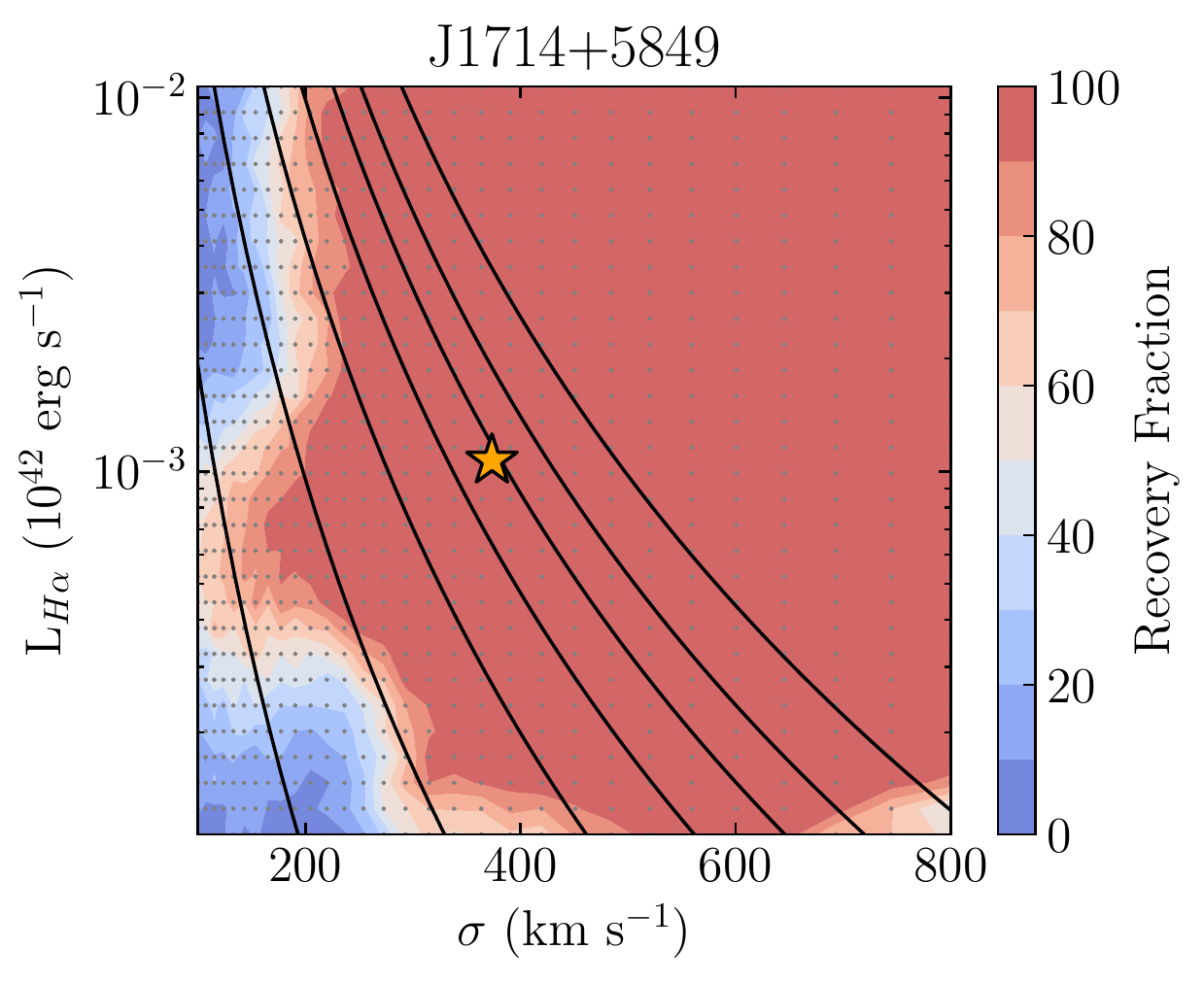} 
\includegraphics[clip,trim={1.04cm 0 0cm 0},height=0.199\hsize]{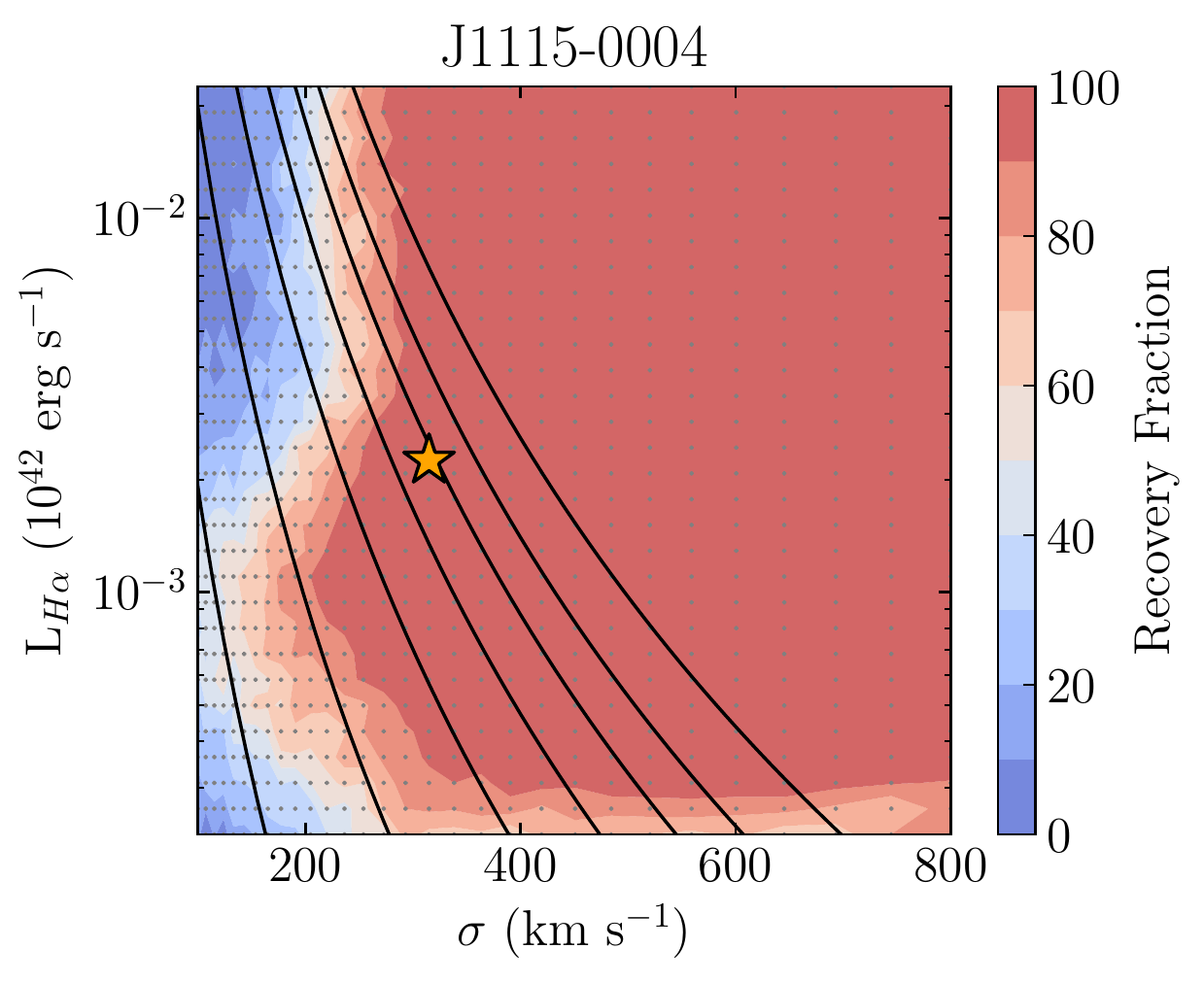} 
}
\centerline{
\includegraphics[clip,trim={0.0cm 0 2.25cm 0},height=0.199\hsize]{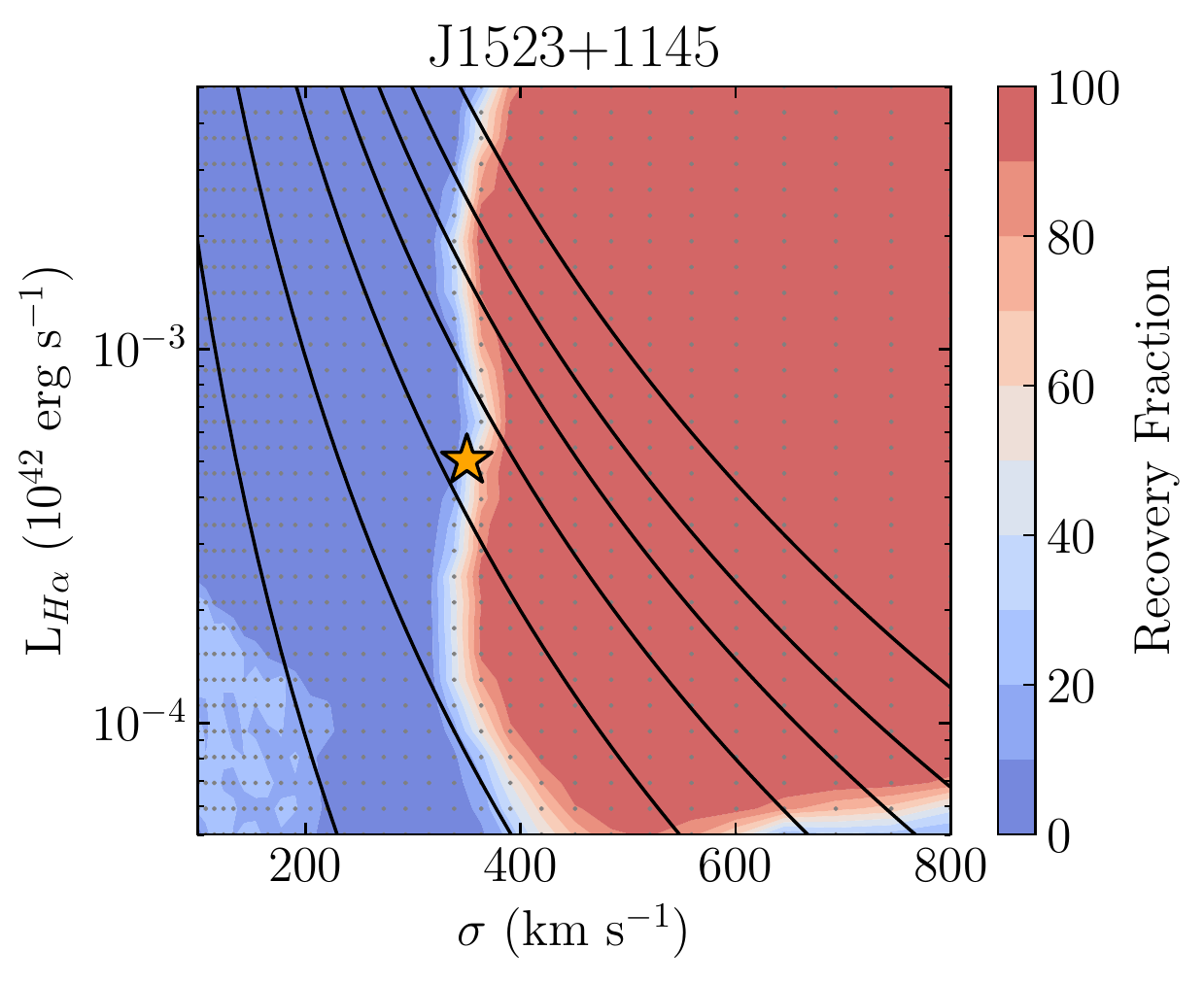} 
\includegraphics[clip,trim={1.04cm 0 2.25cm 0},height=0.199\hsize]{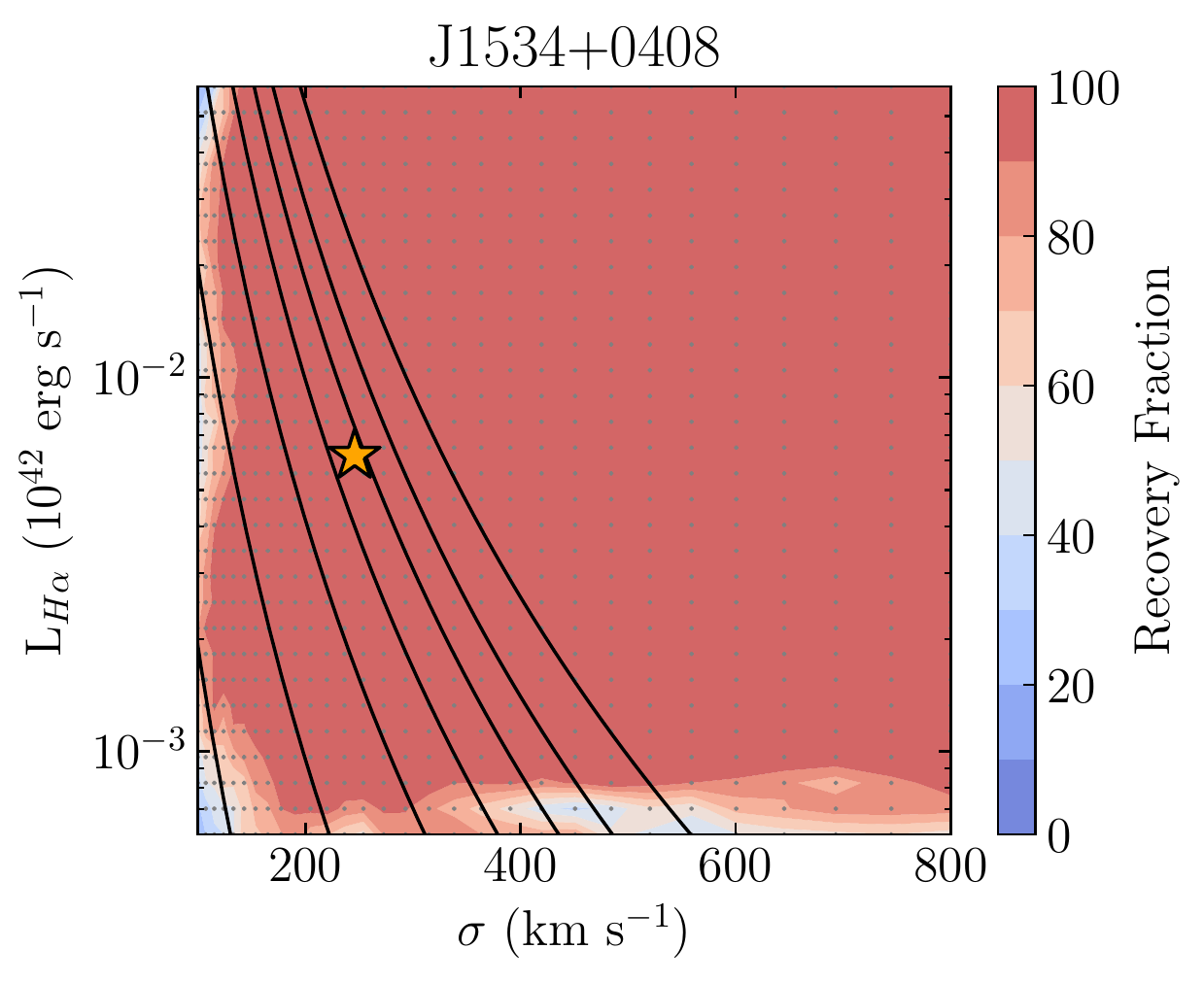} 
\includegraphics[clip,trim={1.04cm 0 2.25cm 0},height=0.199\hsize]{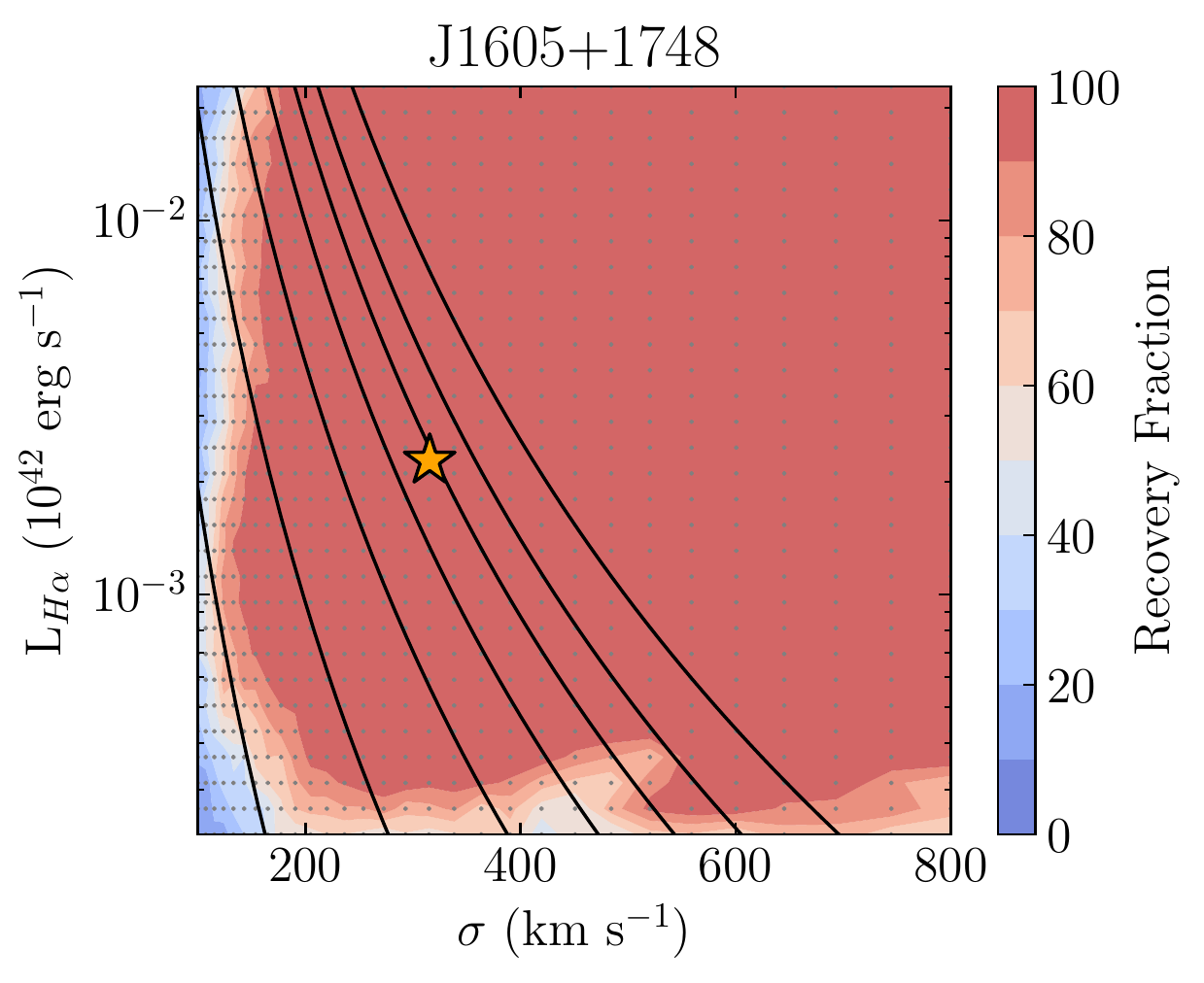} 
\includegraphics[clip,trim={1.04cm 0 2.25cm 0},height=0.199\hsize]{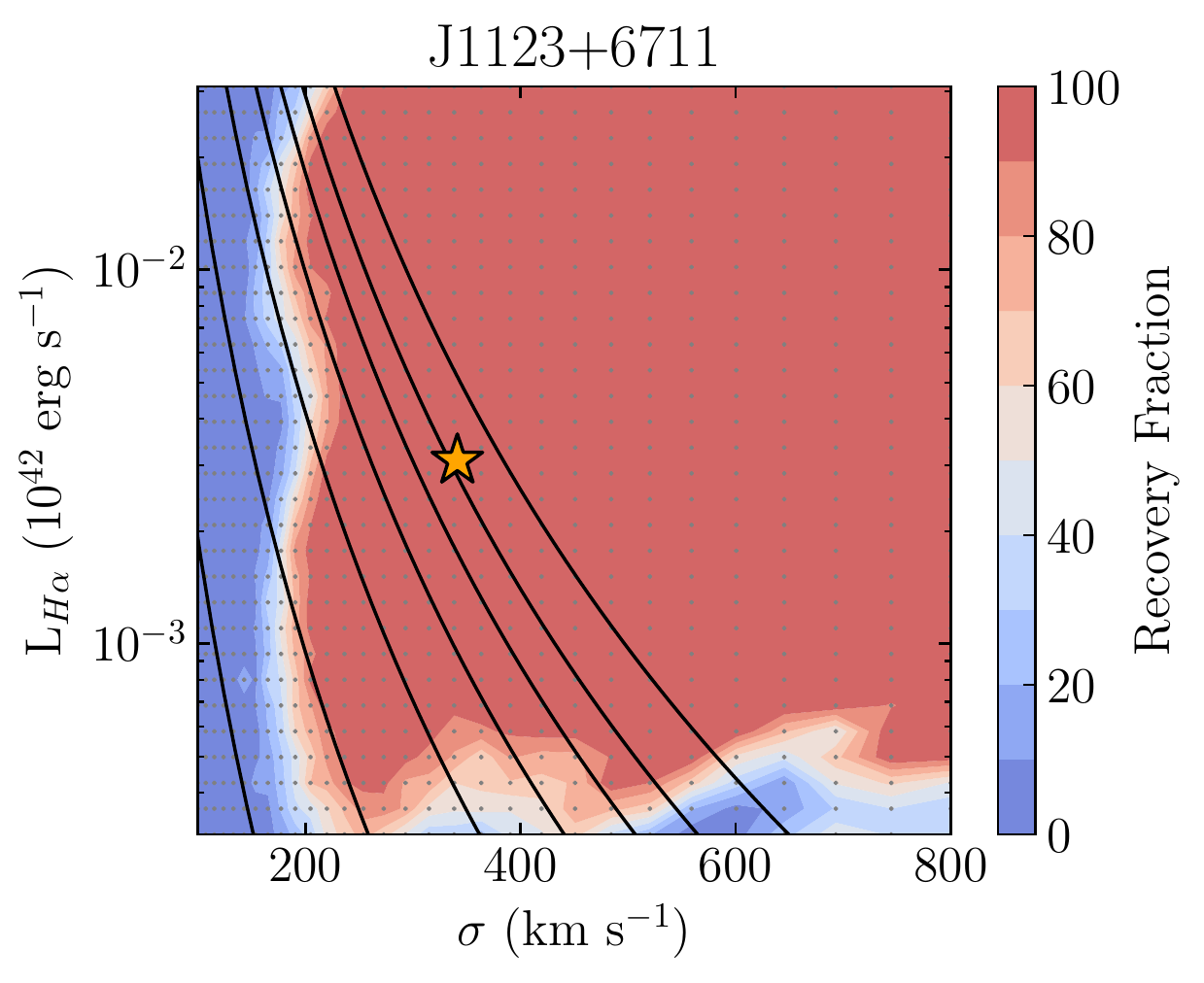} 
\includegraphics[clip,trim={1.04cm 0 0cm 0},height=0.199\hsize]{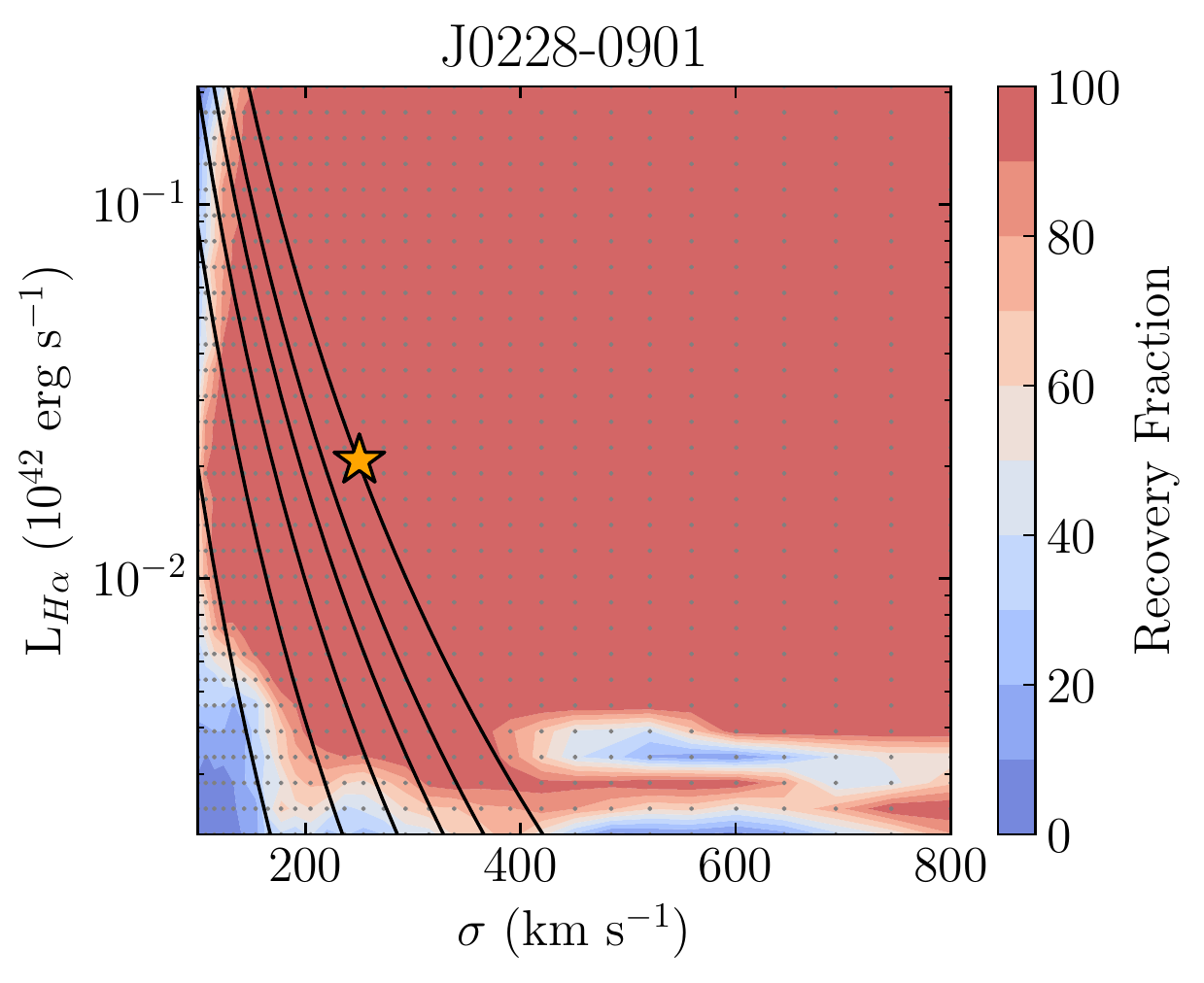} 
}
\caption{Monte-Carlo simulations of the emission line analysis. Black lines show equal black hole masses of $10^4$, $3\times10^4$, $6\times10^4$, $9\times10^4$, $1.2\times10^5$, $1.5\times10^5$, $2\times10^5$ solar masses. Color indicates the recovery fraction of the black hole mass determination in every point of the grid derived from 100 random noise realizations. The source of the spectral data is MagE if indicated on top of the panel or SDSS otherwise. The panels are in the same order as in Fig.~\ref{fig_sample}.\label{fig_mc}}
\end{figure*}

We studied the behavior of our non-parametric emission line analysis algorithm with Monte-Carlo simulations.
For each object in our final sample of 10 X-ray confirmed IMBHs we generated a set of 90,000 mock emission spectra, which included all strong optical emission lines (H$\beta$, [O{\sc iii}], [N{\sc ii}], H$\alpha$, [S{\sc II}]).
To generate synthetic forbidden lines in these spectra we took their model profiles derived from the narrow line LOSVD, which was recovered at the first pass of the algorithm, and added random noise with the distribution derived from the observed spectrum noise in the vicinity of a line.
For allowed lines we also added broad Gaussian components with a random Balmer decrement in the range 2.8--3.2, whose widths $\sigma$ and luminosities $L_{\rm H\alpha}$ were distributed in a grid to cover the region of interest in the parameter space.
At each point of this grid we generated 100 spectra with random noise realizations which were fed to the non-parametric emission line analysis algorithm.
$M_{\rm BH}$ recovered by the algorithm was then compared to the true input value used.
We considered an individual trial successful in case the recovered black hole mass fell within 0.3~dex from the input value, and computed recovery rate at each point of ($\sigma$, $L_{\rm H\alpha}$) grid as a fraction of successful to total number of trials.

The maps of recovery rate for MagE and SDSS spectra of objects from the final sample of 10 IMBHs are presented in Fig.~\ref{fig_mc}.
In almost all cases the derived black holes masses lie in the region with reliable recovery rate.
This modeling also shows that under favorable circumstances our non-parametric emission line analysis algorithm could recover from SDSS spectra black hole masses as low as $10^4 M_{\odot}$.

\subsection{Contamination estimate of the parent sample of IMBH candidates}

We estimate the contamination of the parent sample of IMBH candidates produced by our method using several approaches.
By contamination here we mean the fraction of sources in the sample that are not actual IMBHs, that is they do not exhibit all the required observed IMBH properties: persistent broad H$\alpha$ emission, X-ray emission from an accretion disk and AGN or composite emission lines ratio in the BPT diagram.
By construction, our parent sample contains sources with single-epoch spectroscopic mass estimate without X-ray confirmation for most of them.
Hence, it inevitably contains sources that e.g. showed IMBH features at some point in time and then changed their appearance.
The contaminating sources likely have diverse origin.
These could be supernovae, transient stellar processes \citep{reines13,baldassare16} or spurious detections caused by the imperfection of our spectral analysis.

It is generally very hard to precisely estimate the contamination, so here for simplicity we derive an upper limit of the contamination, i.e. the most pessimistic estimate of the quality of our IMBH search method.
It requires that {\it bona fide} IMBH candidates satisfy the most stringent observing constraints: they must have an X-ray detection and consistent multi-epoch spectroscopic mass estimates (more precisely, we require that broad H$\alpha$ emission is detected at different epochs possibly with different instruments and all its detections satisfy our quality criteria, the mass estimates at different epochs are consistent within 0.3~dex, and the BPT classification does not change).

First we checked our parent sample against the 3XMM-DR5 upper limit server (\url{http://www.ledas.ac.uk/flix/flix.html}) and found 14 objects that serendipitously fell in the footprint of archival {\it XMM-Newton} observations but were not detected in them.
We compared detection limits of these observations with the fluxes expected from the 14 IMBH candidates given existing $L_{\rm X}$---$L_{\rm [OIII]}$ correlation \citep{heckman05}.
None of these observations were deep enough to exclude X-ray emission at the level of $L_{\rm X}$---$L_{\rm [OIII]}$ correlation minus its $1\sigma$ scatter.
We, therefore, cannot reject the accreting IMBH hypothesis for these objects.
Given that the objects of our interest are all low-mass AGN with small luminosities, we expect that other X-ray archives are unlikely to contain many deep enough observations of our IMBH candidates.

Hence, out of 305 IMBH candidates in our parent sample only 18 possess sufficiently deep X-ray observations to confirm or rule out the accreting IMBH hypothesis: 14 from X-ray archives and 4 IMBH candidates observed with dedicated X-ray observations by {\it Chandra} and {\it XMM-Newton} in this work.
Out of those 18, two sources (SDSS~J161251.77+110621.6 observed by {\it XMM-Newton} and SDSS~J135750.71+223100.8 observed by {\it Chandra}) were not detected in X-ray at the level below of that expected from $L_{\rm X}$---$L_{\rm [OIII]}$ correlation minus its $1\sigma$ scatter which secures their non-IMBH nature.
One source, SDSS~J144005.82+115508.7, while detected in X-ray with {\it XMM-Newton} in our observation, did not display broad H$\alpha$ emission at the second spectroscopic epoch when observed with Magellan/MagE.
In addition to this, we discarded five sources without performing the second epoch spectroscopic follow-up observations considering them spurious detections.
This happens, for example, when our automated emission line decomposition algorithm underestimates the broad H$\alpha$ emission flux and after more careful emission line decomposition with manually adjusted constraints, a black hole mass estimate exceeds $2 \times 10^5 M_{\odot}$.
Thus, out of 18 sources with deep X-ray data we discard eight sources in total.
10 remaining sources are those presented in Table~\ref{tableimbh}.
For six sources from Table~\ref{tableimbh} we have both X-ray and second epoch spectroscopic confirmation (objects found in this work: SDSS J1227+0757, SDSS J1342+0530, SDSS J1115$-$0004, SDSS J1107+1347; previously known objects: SDSS J1523+1145, SDSS J0228$-$0901).
Four remaining sources from Table~\ref{tableimbh} (object found in this work: SDSS J1714+5849; previously known objects: SDSS J1534+0408, SDSS J1605+1748, SDSS J1123+6711) are detected in X-ray and have reliable single epoch detections of broad H$\alpha$ emission but do not possess second epoch spectroscopic observations and therefore cannot be used for the contamination estimate.

This leaves us with a sample of 14 sources that have enough evidence to tell if they pass all required tests (multi-epoch spectroscopy and deep enough X-ray observations) or not: 6 sources successfully pass them all, and 8 sources fail in at least one test.
The upper limit on the contamination of our parent sample can be estimated as 8 / 14 = 57\%.
A more realistic estimate should lower this value.
In particular, it was shown \citep{baldassare16} that 100\% of objects classified as AGN on the BPT diagram which at the same time show the broad H$\alpha$ emission, exhibit the same properties in second spectroscopic epoch.
Out of 4 objects without the second spectroscopic epoch, one (SDSS J1605+1748, \citet{dong07}) is classified as AGN on the BPT diagram.
It is then natural to anticipate that it is true IMBH which would lower the contaminating fraction in our sample to 53\%.

Therefore, we have all evidence to expect that at least $0.43 \times 305 = 131$ sources in our parent sample of IMBH candidates are real IMBHs in a sense that they will satisfy the most stringent observing criteria once the necessary follow-up observations have been performed.
If we assume no strong dependence of the contamination level on the black hole mass, we find that at least 42 of our IMBH candidates from Table~\ref{table_all_objs} with $M_{\rm BH} < 10^5 M_{\odot}$ must be actual IMBHs.

\subsection{Implications for co-evolution of central black holes and their host galaxies}

\begin{figure*}
\centering
\includegraphics[width=0.7\hsize]{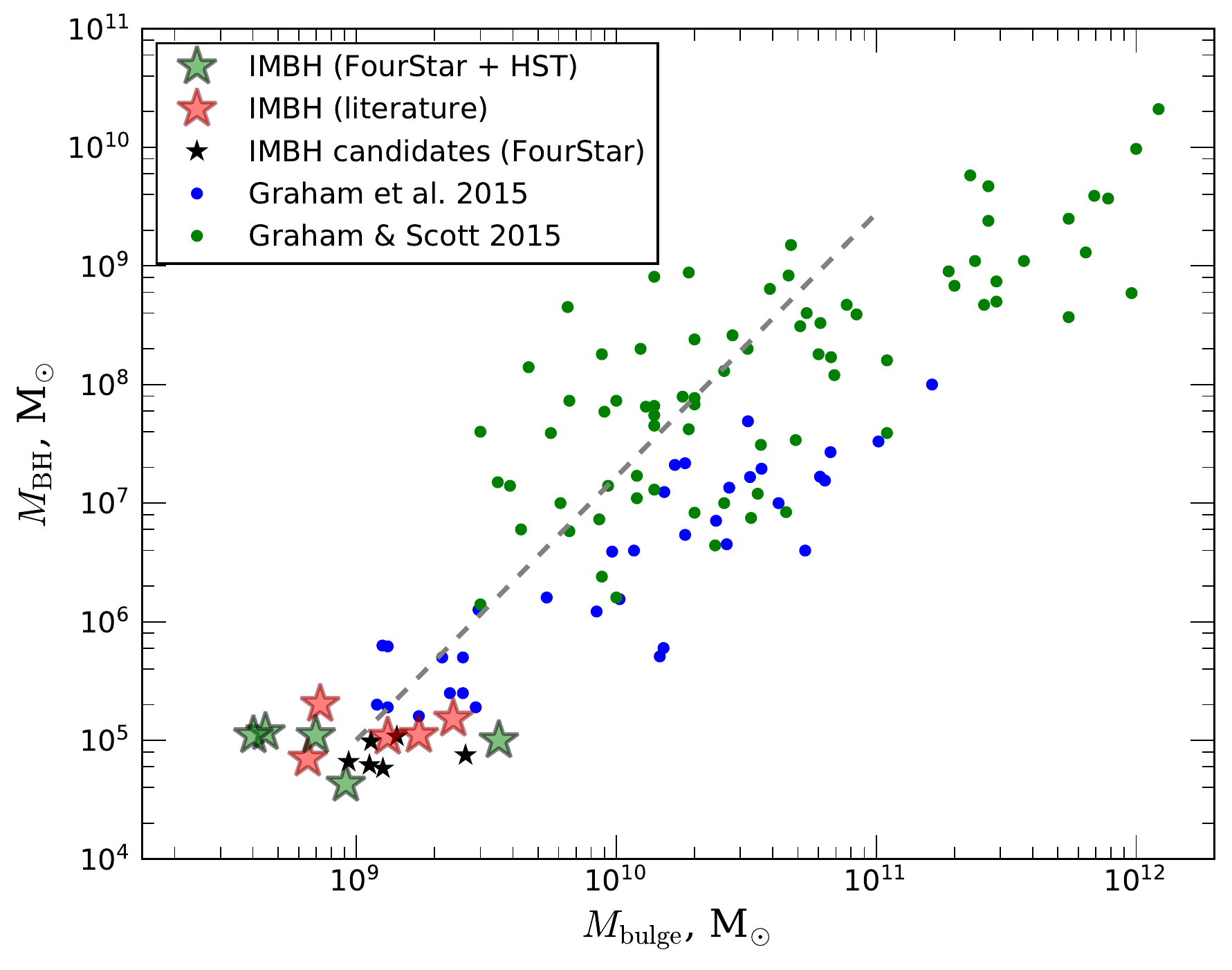}
\caption{A scaling relation between the central black hole mass
$M_{\mathrm{BH}}$ and the mass of a host galaxy bulge/spheroid $M_{\mathrm{bulge}}$. Masses of quiescent and active black holes and stellar masses of bulges of their host galaxies \citep{GS15,graham15} show a strong correlation for galaxies having different morphological types, which supports the scenario of their co-evolution \citep{kormendy13}. The gray dashed line is a power-law fit of the relation for host galaxies with S\'ersic light profiles \citep{GS15}. The X-ray confirmed IMBHs and their hosts (see Fig.~\ref{fig_sample} for their images) are shown as large green (\emph{HST} and \emph{FourStar} observations) and red (literature) star symbols; IMBH candidates without X-ray confirmation (\emph{FourStar} observations) as small black stars. They extend the correlation to lower masses: its continuity suggests that the nuclear IMBHs represent the low mass extension of the mass function of central black holes in galaxies.
\label{fig_screl}}
\end{figure*}

In Fig.~\ref{fig_screl} we compare IMBH masses and host galaxy properties to the recent compilations of bulge/spheroid masses of host galaxies of massive black holes \citep{GS15,graham15}. All IMBH hosts have stellar masses of their bulges between $4\times10^8$ and $4\times10^9 M_{\odot}$ and reside on the low-mass extension of the $M_{\rm BH} - M_{\rm bulge}$ scaling relation established by more massive black holes and their host galaxies, thus filling a sparsely populated region of the parameter space. This argues for the validity of the search approach that looks for AGN signatures created by IMBHs and also supports the connection between the black hole mass growth and the growth of their host galaxy bulges via mergers down to the IMBH regime. 

Galaxy mergers were frequent when the Universe was younger (redshifts $z>1$, \citealp{CBDP03,Bell+06,Lotz+11}). They are thought to be responsible for the growth of bulges \citep{ABP01,BMQ06}, hence suggesting the co-evolution of central black holes and their hosts \citep{kormendy13}, and explaining the observed correlations between central black hole masses and bulge properties, stellar velocity dispersion \citep{FM00,Gebhardt+00,vandenBosch16} and stellar mass \citep{MH03,HR04,Gultekin+09}. Therefore, IMBH host galaxies must have experienced very few if any major mergers over their lifetime.

From the multiple epoch optical spectroscopy and X-ray observations, we estimate that our IMBH candidate sample may include up-to 57\%\ of transient broad lines or spurious detections (see a detailed discussion above). Even though, keeping in mind that virial masses are uncertain to a factor of 2, it should contain at least 42 objects with masses smaller than $10^5 M_{\odot}$. These objects are the relics of the early SMBH formation survived through the cosmic time almost intact and their host galaxies must have had very poor merger histories. Their existence suggests that at least some SMBHs did not originate from massive ($>10^5 M_{\odot}$) seeds but from stellar mass black holes. The efficiency of mass growth via super-Eddington accretion is questionable because of radiation driven outflows \citep{king16}. Therefore, stellar mass black hole mergers must have played an important role in the SMBH assembly.

Our sample likely represents the tip of the iceberg of the IMBH population. The sphere of influence of an IMBH is too small to significantly affect the stellar dynamics of its host galaxy and cannot be detected beyond the Local Group with currently available instruments, therefore we can find only IMBHs in the actively accreting phase, which requires the gas supply onto the galaxy centre. On the other hand, most non-starforming galaxies with small bulges reside in galaxy clusters, which does not favour AGN, because they lost their gas completely due to environmental effects. Therefore, while the exact fraction of actively accreting IMBHs is unknown, it is likely smaller than that of more massive black holes. 

The main limitations of our technique are the relatively shallow flux limit of the SDSS spectroscopy and the lack of wide field X-ray surveys reaching the flux limit ($5\times10^{-15}$~erg~cm$^{-2}$~s$^{-1}$) of a snapshot \emph{Chandra} or \emph{XMM-Newton} observation: 4 serendipitously detected sources reside in $\simeq2$~\%\ of the sky observed by both SDSS and X-ray satellites. The future deep \emph{eRosita} X-ray survey may confirm several dozen IMBH candidates from our current sample of 305. A targeted optical spectroscopic probe of nearby galaxies with small bulges deeper than SDSS will likely bring new IMBH identifications at even lower masses.

\acknowledgments
IC, IK, IZ, and KG acknowledge the support by the Russian
Scientific Foundation grant 17-72-20119 that was used in the final stages of this work and the manuscript preparation.
IZ acknowledges the Russian Scientific Foundation grant 14-50-00043 for the development of the pipeline system used for input catalog assembly.
The authors acknowledge partial support from the M.V.Lomonosov
Moscow State University Program of Development, and a Russian--French PICS International Laboratory program (no.  6590) co-funded by the RFBR (project 15-52-15050), entitled ``Galaxy evolution mechanisms in the Local Universe and at intermediate redshifts''.
The authors are grateful to citizen scientist A.~Zolotov for his help with the figure 1 of the manuscript.
Support for this work was provided by the National Aeronautics and Space Administration through Chandra Award Number 18708581 issued by the Chandra X-ray Center, which is operated by the Smithsonian Astrophysical Observatory for and on behalf of the National Aeronautics Space Administration under contract NAS8-03060.
The scientific results reported in this article are based in part on observations made by the Chandra X-ray Observatory (observations 20114, 20115).
Based on observations obtained with {\it XMM-Newton}, an ESA science mission with instruments and contributions directly funded by ESA Member States and NASA (observations 0795711301, 0795711401).
This paper includes data gathered with the 6.5 meter Magellan Telescopes located at Las Campanas Observatory, Chile.
Based on observations made with the NASA/ESA Hubble Space Telescope, obtained from the Data Archive at the Space Telescope Science Institute, which is operated by the Association of Universities for Research in Astronomy, Inc., under NASA contract NAS 5-26555.
These observations are associated with the {\it HST} program 11142.
This research has made use of TOPCAT, developed by Mark Taylor at the University of Bristol; Aladin developed by the Centre de Donn\'ees Astronomiques de Strasbourg (CDS); the VizieR catalogue access tool (CDS); Astropy, a community-developed core Python package for Astronomy (Astropy Collaboration, 2013); Atlassian JIRA issue tracking system and Bitbucket source code hosting service.
Funding for the \emph{SDSS} and \emph{SDSS}-II has been provided by the Alfred P.  Sloan Foundation, the Participating Institutions, the National Science Foundation, the U.S. Department of Energy, the National Aeronautics and Space Administration, the Japanese Monbukagakusho, the Max Planck Society, and the Higher Education Funding Council for England.
The \emph{SDSS} Web Site is \url{http://www.sdss.org/}. 
The Pan-STARRS1 Surveys (PS1) and the PS1 public science archive have been made possible through contributions by the Institute for Astronomy, the University of Hawaii, the Pan-STARRS Project Office, the Max-Planck Society and its participating institutes, the Max Planck Institute for Astronomy, Heidelberg and the Max Planck Institute for Extraterrestrial Physics, Garching, The Johns Hopkins University, Durham University, the University of Edinburgh, the Queen's University Belfast, the Harvard-Smithsonian Center for Astrophysics, the Las Cumbres Observatory Global Telescope Network Incorporated, the National Central University of Taiwan, the Space Telescope Science Institute, the National Aeronautics and Space Administration under Grant No. NNX08AR22G issued through the Planetary Science Division of the NASA Science Mission Directorate, the National Science Foundation Grant No. AST-1238877, the University of Maryland, Eotvos Lorand University (ELTE), the Los Alamos National Laboratory, and the Gordon and Betty Moore Foundation.

\bibliographystyle{aasjournal}
\bibliography{IMBH}

\begin{thebibliography}{}
\expandafter\ifx\csname natexlab\endcsname\relax\def\natexlab#1{#1}\fi
\providecommand{\url}[1]{\href{#1}{#1}}

\bibitem[{{Abazajian} {et~al.}(2009){Abazajian}, {Adelman-McCarthy},
  {Ag{\"u}eros}, {Allam}, {Allende Prieto}, {An}, {Anderson}, {Anderson},
  {Annis}, {Bahcall}, \& et~al.}]{SDSS_DR7}
{Abazajian}, K.~N., {Adelman-McCarthy}, J.~K., {Ag{\"u}eros}, M.~A., {et~al.}
  2009, \apjs, 182, 543

\bibitem[{{Abbott} {et~al.}(2016){Abbott}, {Abbott}, {Abbott}, {Abernathy},
  {Acernese}, {Ackley}, {Adams}, {Adams}, {Addesso}, {Adhikari}, \&
  et~al.}]{LIGO1}
{Abbott}, B.~P., {Abbott}, R., {Abbott}, T.~D., {et~al.} 2016, Physical Review
  Letters, 116, 061102

\bibitem[{{Aguerri} {et~al.}(2001){Aguerri}, {Balcells}, \& {Peletier}}]{ABP01}
{Aguerri}, J.~A.~L., {Balcells}, M., \& {Peletier}, R.~F. 2001, \aap, 367, 428

\bibitem[{{Baldassare} {et~al.}(2015){Baldassare}, {Reines}, {Gallo}, \&
  {Greene}}]{baldassare15}
{Baldassare}, V.~F., {Reines}, A.~E., {Gallo}, E., \& {Greene}, J.~E. 2015,
  \apjl, 809, L14

\bibitem[{{Baldassare} {et~al.}(2016){Baldassare}, {Reines}, {Gallo}, {Greene},
  {Graur}, {Geha}, {Hainline}, {Carroll}, \& {Hickox}}]{baldassare16}
{Baldassare}, V.~F., {Reines}, A.~E., {Gallo}, E., {et~al.} 2016, \apj, 829, 57

\bibitem[{{Baldwin} {et~al.}(1981){Baldwin}, {Phillips}, \&
  {Terlevich}}]{BPT81}
{Baldwin}, J.~A., {Phillips}, M.~M., \& {Terlevich}, R. 1981, \pasp, 93, 5

\bibitem[{{Barth} {et~al.}(2004){Barth}, {Ho}, {Rutledge}, \&
  {Sargent}}]{BHRS04}
{Barth}, A.~J., {Ho}, L.~C., {Rutledge}, R.~E., \& {Sargent}, W.~L.~W. 2004,
  \apj, 607, 90

\bibitem[{{Begelman} {et~al.}(2006){Begelman}, {Volonteri}, \&
  {Rees}}]{begelman06}
{Begelman}, M.~C., {Volonteri}, M., \& {Rees}, M.~J. 2006, \mnras, 370, 289

\bibitem[{{Bell} {et~al.}(2006){Bell}, {Phleps}, {Somerville}, {Wolf}, {Borch},
  \& {Meisenheimer}}]{Bell+06}
{Bell}, E.~F., {Phleps}, S., {Somerville}, R.~S., {et~al.} 2006, \apj, 652, 270

\bibitem[{{Bertin}(2011)}]{bertin11}
{Bertin}, E. 2011, in Astronomical Society of the Pacific Conference Series,
  Vol. 442, Astronomical Data Analysis Software and Systems XX, ed. I.~N.
  {Evans}, A.~{Accomazzi}, D.~J. {Mink}, \& A.~H. {Rots}, 435

\bibitem[{{Boylan-Kolchin} {et~al.}(2006){Boylan-Kolchin}, {Ma}, \&
  {Quataert}}]{BMQ06}
{Boylan-Kolchin}, M., {Ma}, C.-P., \& {Quataert}, E. 2006, \mnras, 369, 1081

\bibitem[{{Chambers} {et~al.}(2016){Chambers}, {Magnier}, {Metcalfe},
  {Flewelling}, {Huber}, {Waters}, {Denneau}, {Draper}, {Farrow}, {Finkbeiner},
  {Holmberg}, {Koppenhoefer}, {Price}, {Saglia}, {Schlafly}, {Smartt},
  {Sweeney}, {Wainscoat}, {Burgett}, {Grav}, {Heasley}, {Hodapp}, {Jedicke},
  {Kaiser}, {Kudritzki}, {Luppino}, {Lupton}, {Monet}, {Morgan}, {Onaka},
  {Stubbs}, {Tonry}, {Banados}, {Bell}, {Bender}, {Bernard}, {Botticella},
  {Casertano}, {Chastel}, {Chen}, {Chen}, {Cole}, {Deacon}, {Frenk},
  {Fitzsimmons}, {Gezari}, {Goessl}, {Goggia}, {Goldman}, {Grebel}, {Hambly},
  {Hasinger}, {Heavens}, {Heckman}, {Henderson}, {Henning}, {Holman}, {Hopp},
  {Ip}, {Isani}, {Keyes}, {Koekemoer}, {Kotak}, {Long}, {Lucey}, {Liu},
  {Martin}, {McLean}, {Morganson}, {Murphy}, {Nieto-Santisteban}, {Norberg},
  {Peacock}, {Pier}, {Postman}, {Primak}, {Rae}, {Rest}, {Riess}, {Riffeser},
  {Rix}, {Roser}, {Schilbach}, {Schultz}, {Scolnic}, {Szalay}, {Seitz},
  {Shiao}, {Small}, {Smith}, {Soderblom}, {Taylor}, {Thakar}, {Thiel},
  {Thilker}, {Urata}, {Valenti}, {Walter}, {Watters}, {Werner}, {White},
  {Wood-Vasey}, \& {Wyse}}]{2016arXiv161205560C}
{Chambers}, K.~C., {Magnier}, E.~A., {Metcalfe}, N., {et~al.} 2016, ArXiv
  e-prints, arXiv:1612.05560

\bibitem[{{Chilingarian} {et~al.}(2007{\natexlab{a}}){Chilingarian},
  {Prugniel}, {Sil'Chenko}, \& {Koleva}}]{Chilingarian07a}
{Chilingarian}, I., {Prugniel}, P., {Sil'Chenko}, O., \& {Koleva}, M.
  2007{\natexlab{a}}, in IAU Symposium, Vol. 241, Stellar Populations as
  Building Blocks of Galaxies, ed. A.~{Vazdekis} \& R.~{Peletier}, 175--176

\bibitem[{{Chilingarian} {et~al.}(2007{\natexlab{b}}){Chilingarian},
  {Prugniel}, {Sil'Chenko}, \& {Afanasiev}}]{CPSA07}
{Chilingarian}, I.~V., {Prugniel}, P., {Sil'Chenko}, O.~K., \& {Afanasiev},
  V.~L. 2007{\natexlab{b}}, \mnras, 376, 1033

\bibitem[{{Chilingarian} {et~al.}(2017){Chilingarian}, {Zolotukhin}, {Katkov},
  {Melchior}, {Rubtsov}, \& {Grishin}}]{RCSED}
{Chilingarian}, I.~V., {Zolotukhin}, I.~Y., {Katkov}, I.~Y., {et~al.} 2017,
  \apjs, 228, 14

\bibitem[{{Conselice} {et~al.}(2003){Conselice}, {Bershady}, {Dickinson}, \&
  {Papovich}}]{CBDP03}
{Conselice}, C.~J., {Bershady}, M.~A., {Dickinson}, M., \& {Papovich}, C. 2003,
  \aj, 126, 1183

\bibitem[{{Dong} {et~al.}(2012){Dong}, {Greene}, \& {Ho}}]{dong12}
{Dong}, R., {Greene}, J.~E., \& {Ho}, L.~C. 2012, \apj, 761, 73

\bibitem[{{Dong} {et~al.}(2007){Dong}, {Wang}, {Yuan}, {Shan}, {Zhou}, {Fan},
  {Dou}, {Wang}, {Wang}, \& {Lu}}]{dong07}
{Dong}, X., {Wang}, T., {Yuan}, W., {et~al.} 2007, \apj, 657, 700

\bibitem[{{Dou} {et~al.}(2016){Dou}, {Wang}, {Jiang}, {Yang}, {Lyu}, \&
  {Zhou}}]{dou16}
{Dou}, L., {Wang}, T.-g., {Jiang}, N., {et~al.} 2016, \apj, 832, 188

\bibitem[{{Elvis}(2000)}]{elvis00}
{Elvis}, M. 2000, \apj, 545, 63

\bibitem[{{Evans} {et~al.}(2010){Evans}, {Primini}, {Glotfelty}, {Anderson},
  {Bonaventura}, {Chen}, {Davis}, {Doe}, {Evans}, {Fabbiano}, {Galle}, {Gibbs},
  {Grier}, {Hain}, {Hall}, {Harbo}, {(Helen He}, {Houck}, {Karovska},
  {Kashyap}, {Lauer}, {McCollough}, {McDowell}, {Miller}, {Mitschang},
  {Morgan}, {Mossman}, {Nichols}, {Nowak}, {Plummer}, {Refsdal}, {Rots},
  {Siemiginowska}, {Sundheim}, {Tibbetts}, {Van Stone}, {Winkelman}, \&
  {Zografou}}]{evans10}
{Evans}, I.~N., {Primini}, F.~A., {Glotfelty}, K.~J., {et~al.} 2010, \apjs,
  189, 37

\bibitem[{{Evans} {et~al.}(2014){Evans}, {Osborne}, {Beardmore}, {Page},
  {Willingale}, {Mountford}, {Pagani}, {Burrows}, {Kennea}, {Perri},
  {Tagliaferri}, \& {Gehrels}}]{evans14}
{Evans}, P.~A., {Osborne}, J.~P., {Beardmore}, A.~P., {et~al.} 2014, \apjs,
  210, 8

\bibitem[{{Ferrarese} \& {Merritt}(2000)}]{FM00}
{Ferrarese}, L., \& {Merritt}, D. 2000, \apjl, 539, L9

\bibitem[{{Filippenko} \& {Ho}(2003)}]{FH03}
{Filippenko}, A.~V., \& {Ho}, L.~C. 2003, \apjl, 588, L13

\bibitem[{{Filippenko} \& {Sargent}(1989)}]{FS89}
{Filippenko}, A.~V., \& {Sargent}, W.~L.~W. 1989, \apjl, 342, L11

\bibitem[{{Gebhardt} {et~al.}(2000){Gebhardt}, {Bender}, {Bower}, {Dressler},
  {Faber}, {Filippenko}, {Green}, {Grillmair}, {Ho}, {Kormendy}, {Lauer},
  {Magorrian}, {Pinkney}, {Richstone}, \& {Tremaine}}]{Gebhardt+00}
{Gebhardt}, K., {Bender}, R., {Bower}, G., {et~al.} 2000, \apjl, 539, L13

\bibitem[{{Graham} {et~al.}(2015){Graham}, {Dullo}, \& {Savorgnan}}]{graham15}
{Graham}, A.~W., {Dullo}, B.~T., \& {Savorgnan}, G.~A.~D. 2015, \apj, 804, 32

\bibitem[{{Graham} \& {Scott}(2015)}]{GS15}
{Graham}, A.~W., \& {Scott}, N. 2015, \apj, 798, 54

\bibitem[{{Greene} \& {Ho}(2005)}]{greene05}
{Greene}, J.~E., \& {Ho}, L.~C. 2005, \apj, 630, 122

\bibitem[{{Greene} \& {Ho}(2007)}]{2007ApJ...670...92G}
---. 2007, \apj, 670, 92

\bibitem[{{G{\"u}ltekin} {et~al.}(2009){G{\"u}ltekin}, {Richstone}, {Gebhardt},
  {Lauer}, {Tremaine}, {Aller}, {Bender}, {Dressler}, {Faber}, {Filippenko},
  {Green}, {Ho}, {Kormendy}, {Magorrian}, {Pinkney}, \& {Siopis}}]{Gultekin+09}
{G{\"u}ltekin}, K., {Richstone}, D.~O., {Gebhardt}, K., {et~al.} 2009, \apj,
  698, 198

\bibitem[{{H{\"a}ring} \& {Rix}(2004)}]{HR04}
{H{\"a}ring}, N., \& {Rix}, H.-W. 2004, \apjl, 604, L89

\bibitem[{{Heckman} {et~al.}(2005{\natexlab{a}}){Heckman}, {Ptak},
  {Hornschemeier}, \& {Kauffmann}}]{2005ApJ...634..161H}
{Heckman}, T.~M., {Ptak}, A., {Hornschemeier}, A., \& {Kauffmann}, G.
  2005{\natexlab{a}}, \apj, 634, 161

\bibitem[{{Heckman} {et~al.}(2005{\natexlab{b}}){Heckman}, {Ptak},
  {Hornschemeier}, \& {Kauffmann}}]{heckman05}
---. 2005{\natexlab{b}}, \apj, 634, 161

\bibitem[{{Hinshaw} {et~al.}(2013){Hinshaw}, {Larson}, {Komatsu}, {Spergel},
  {Bennett}, {Dunkley}, {Nolta}, {Halpern}, {Hill}, {Odegard}, {Page}, {Smith},
  {Weiland}, {Gold}, {Jarosik}, {Kogut}, {Limon}, {Meyer}, {Tucker}, {Wollack},
  \& {Wright}}]{hinshaw13}
{Hinshaw}, G., {Larson}, D., {Komatsu}, E., {et~al.} 2013, \apjs, 208, 19

\bibitem[{{Kewley} {et~al.}(2006){Kewley}, {Groves}, {Kauffmann}, \&
  {Heckman}}]{kewley06}
{Kewley}, L.~J., {Groves}, B., {Kauffmann}, G., \& {Heckman}, T. 2006, \mnras,
  372, 961

\bibitem[{{King} \& {Lasota}(2014)}]{king14}
{King}, A., \& {Lasota}, J.-P. 2014, \mnras, 444, L30

\bibitem[{{King} \& {Muldrew}(2016)}]{king16}
{King}, A., \& {Muldrew}, S.~I. 2016, \mnras, 455, 1211

\bibitem[{{K{\i}z{\i}ltan} {et~al.}(2017){K{\i}z{\i}ltan}, {Baumgardt}, \&
  {Loeb}}]{kiziltan17}
{K{\i}z{\i}ltan}, B., {Baumgardt}, H., \& {Loeb}, A. 2017, \nat, 542, 203

\bibitem[{{Kormendy} \& {Ho}(2013)}]{kormendy13}
{Kormendy}, J., \& {Ho}, L.~C. 2013, \araa, 51, 511

\bibitem[{{Kormendy} \& {Richstone}(1995)}]{1995ARA&A..33..581K}
{Kormendy}, J., \& {Richstone}, D. 1995, \araa, 33, 581

\bibitem[{{Kunth} {et~al.}(1987){Kunth}, {Sargent}, \& {Bothun}}]{KSB87}
{Kunth}, D., {Sargent}, W.~L.~W., \& {Bothun}, G.~D. 1987, \aj, 93, 29

\bibitem[{{Loeb} \& {Rasio}(1994)}]{loeb94}
{Loeb}, A., \& {Rasio}, F.~A. 1994, \apj, 432, 52

\bibitem[{{Lotz} {et~al.}(2011){Lotz}, {Jonsson}, {Cox}, {Croton}, {Primack},
  {Somerville}, \& {Stewart}}]{Lotz+11}
{Lotz}, J.~M., {Jonsson}, P., {Cox}, T.~J., {et~al.} 2011, \apj, 742, 103

\bibitem[{{Madau} \& {Rees}(2001)}]{madau01}
{Madau}, P., \& {Rees}, M.~J. 2001, \apjl, 551, L27

\bibitem[{{Marconi} \& {Hunt}(2003)}]{MH03}
{Marconi}, A., \& {Hunt}, L.~K. 2003, \apjl, 589, L21

\bibitem[{{Marshall} {et~al.}(2008){Marshall}, {Burles}, {Thompson},
  {Shectman}, {Bigelow}, {Burley}, {Birk}, {Estrada}, {Jones}, {Smith},
  {Kowal}, {Castillo}, {Storts}, \& {Ortiz}}]{2008SPIE.7014E..54M}
{Marshall}, J.~L., {Burles}, S., {Thompson}, I.~B., {et~al.} 2008, in
  \procspie, Vol. 7014, Ground-based and Airborne Instrumentation for Astronomy
  II, 701454

\bibitem[{{Merritt} \& {Milosavljevi{\'c}}(2005)}]{2005LRR.....8....8M}
{Merritt}, D., \& {Milosavljevi{\'c}}, M. 2005, Living Reviews in Relativity,
  8, astro-ph/0410364

\bibitem[{{Miyoshi} {et~al.}(1995){Miyoshi}, {Moran}, {Herrnstein},
  {Greenhill}, {Nakai}, {Diamond}, \& {Inoue}}]{miyoshi95}
{Miyoshi}, M., {Moran}, J., {Herrnstein}, J., {et~al.} 1995, \nat, 373, 127

\bibitem[{{Mortlock} {et~al.}(2011){Mortlock}, {Warren}, {Venemans}, {Patel},
  {Hewett}, {McMahon}, {Simpson}, {Theuns}, {Gonz{\'a}les-Solares}, {Adamson},
  {Dye}, {Hambly}, {Hirst}, {Irwin}, {Kuiper}, {Lawrence}, \&
  {R{\"o}ttgering}}]{mortlock11}
{Mortlock}, D.~J., {Warren}, S.~J., {Venemans}, B.~P., {et~al.} 2011, \nat,
  474, 616

\bibitem[{{Noyola} {et~al.}(2010){Noyola}, {Gebhardt}, {Kissler-Patig},
  {L{\"u}tzgendorf}, {Jalali}, {de Zeeuw}, \&
  {Baumgardt}}]{2010ApJ...719L..60N}
{Noyola}, E., {Gebhardt}, K., {Kissler-Patig}, M., {et~al.} 2010, \apjl, 719,
  L60

\bibitem[{{Peng} {et~al.}(2010){Peng}, {Ho}, {Impey}, \& {Rix}}]{Peng10}
{Peng}, C.~Y., {Ho}, L.~C., {Impey}, C.~D., \& {Rix}, H.-W. 2010, \aj, 139,
  2097

\bibitem[{{Persson} {et~al.}(2013){Persson}, {Murphy}, {Smee}, {Birk},
  {Monson}, {Uomoto}, {Koch}, {Shectman}, {Barkhouser}, {Orndorff}, {Hammond},
  {Harding}, {Scharfstein}, {Kelson}, {Marshall}, \&
  {McCarthy}}]{2013PASP..125..654P}
{Persson}, S.~E., {Murphy}, D.~C., {Smee}, S., {et~al.} 2013, \pasp, 125, 654

\bibitem[{{Peterson} {et~al.}(2005){Peterson}, {Bentz}, {Desroches},
  {Filippenko}, {Ho}, {Kaspi}, {Laor}, {Maoz}, {Moran}, {Pogge}, \&
  {Quillen}}]{Peterson+05}
{Peterson}, B.~M., {Bentz}, M.~C., {Desroches}, L.-B., {et~al.} 2005, \apj,
  632, 799

\bibitem[{{Portegies Zwart} {et~al.}(2004){Portegies Zwart}, {Baumgardt},
  {Hut}, {Makino}, \& {McMillan}}]{2004Natur.428..724P}
{Portegies Zwart}, S.~F., {Baumgardt}, H., {Hut}, P., {Makino}, J., \&
  {McMillan}, S.~L.~W. 2004, \nat, 428, 724

\bibitem[{{Reines} {et~al.}(2013){Reines}, {Greene}, \& {Geha}}]{reines13}
{Reines}, A.~E., {Greene}, J.~E., \& {Geha}, M. 2013, \apj, 775, 116

\bibitem[{{Rosen} {et~al.}(2016){Rosen}, {Webb}, {Watson}, {Ballet}, {Barret},
  {Braito}, {Carrera}, {Ceballos}, {Coriat}, {Della Ceca}, {Denkinson},
  {Esquej}, {Farrell}, {Freyberg}, {Gris{\'e}}, {Guillout}, {Heil},
  {Koliopanos}, {Law-Green}, {Lamer}, {Lin}, {Martino}, {Michel}, {Motch},
  {Nebot Gomez-Moran}, {Page}, {Page}, {Page}, {Pakull}, {Pye}, {Read},
  {Rodriguez}, {Sakano}, {Saxton}, {Schwope}, {Scott}, {Sturm}, {Traulsen},
  {Yershov}, \& {Zolotukhin}}]{rosen16}
{Rosen}, S.~R., {Webb}, N.~A., {Watson}, M.~G., {et~al.} 2016, \aap, 590, A1

\bibitem[{{Saulder} {et~al.}(2016){Saulder}, {van Kampen}, {Chilingarian},
  {Mieske}, \& {Zeilinger}}]{saulder+16}
{Saulder}, C., {van Kampen}, E., {Chilingarian}, I.~V., {Mieske}, S., \&
  {Zeilinger}, W.~W. 2016, \aap, 596, A14

\bibitem[{{Sch{\"o}del} {et~al.}(2002){Sch{\"o}del}, {Ott}, {Genzel},
  {Hofmann}, {Lehnert}, {Eckart}, {Mouawad}, {Alexander}, {Reid}, {Lenzen},
  {Hartung}, {Lacombe}, {Rouan}, {Gendron}, {Rousset}, {Lagrange}, {Brandner},
  {Ageorges}, {Lidman}, {Moorwood}, {Spyromilio}, {Hubin}, \& {Menten}}]{SgrA}
{Sch{\"o}del}, R., {Ott}, T., {Genzel}, R., {et~al.} 2002, \nat, 419, 694

\bibitem[{{Seth} {et~al.}(2014){Seth}, {van den Bosch}, {Mieske}, {Baumgardt},
  {Brok}, {Strader}, {Neumayer}, {Chilingarian}, {Hilker}, {McDermid},
  {Spitler}, {Brodie}, {Frank}, \& {Walsh}}]{seth14}
{Seth}, A.~C., {van den Bosch}, R., {Mieske}, S., {et~al.} 2014, \nat, 513, 398

\bibitem[{{Simard} {et~al.}(2011){Simard}, {Mendel}, {Patton}, {Ellison}, \&
  {McConnachie}}]{simard11}
{Simard}, L., {Mendel}, J.~T., {Patton}, D.~R., {Ellison}, S.~L., \&
  {McConnachie}, A.~W. 2011, \apjs, 196, 11

\bibitem[{{Skrutskie} {et~al.}(2006){Skrutskie}, {Cutri}, {Stiening},
  {Weinberg}, {Schneider}, {Carpenter}, {Beichman}, {Capps}, {Chester},
  {Elias}, {Huchra}, {Liebert}, {Lonsdale}, {Monet}, {Price}, {Seitzer},
  {Jarrett}, {Kirkpatrick}, {Gizis}, {Howard}, {Evans}, {Fowler}, {Fullmer},
  {Hurt}, {Light}, {Kopan}, {Marsh}, {McCallon}, {Tam}, {Van Dyk}, \&
  {Wheelock}}]{skrutskie06}
{Skrutskie}, M.~F., {Cutri}, R.~M., {Stiening}, R., {et~al.} 2006, \aj, 131,
  1163

\bibitem[{{Thornton} {et~al.}(2008){Thornton}, {Barth}, {Ho}, {Rutledge}, \&
  {Greene}}]{Thornton+08}
{Thornton}, C.~E., {Barth}, A.~J., {Ho}, L.~C., {Rutledge}, R.~E., \& {Greene},
  J.~E. 2008, \apj, 686, 892

\bibitem[{{van den Bosch}(2016)}]{vandenBosch16}
{van den Bosch}, R.~C.~E. 2016, \apj, 831, 134

\bibitem[{{Volonteri}(2012)}]{2012Sci...337..544V}
{Volonteri}, M. 2012, Science, 337, 544

\bibitem[{{Webb} {et~al.}(2012){Webb}, {Cseh}, {Lenc}, {Godet}, {Barret},
  {Corbel}, {Farrell}, {Fender}, {Gehrels}, \& {Heywood}}]{webb12}
{Webb}, N., {Cseh}, D., {Lenc}, E., {et~al.} 2012, Science, 337, 554

\bibitem[{{Wrobel} \& {Ho}(2006)}]{WH06}
{Wrobel}, J.~M., \& {Ho}, L.~C. 2006, \apjl, 646, L95

\bibitem[{{Wu} {et~al.}(2015){Wu}, {Wang}, {Fan}, {Yi}, {Zuo}, {Bian}, {Jiang},
  {McGreer}, {Wang}, {Yang}, {Yang}, {Thompson}, \&
  {Beletsky}}]{2015Natur.518..512W}
{Wu}, X.-B., {Wang}, F., {Fan}, X., {et~al.} 2015, \nat, 518, 512

\bibitem[{{Zocchi} {et~al.}(2017){Zocchi}, {Gieles}, \&
  {H{\'e}nault-Brunet}}]{2017MNRAS.468.4429Z}
{Zocchi}, A., {Gieles}, M., \& {H{\'e}nault-Brunet}, V. 2017, \mnras, 468, 4429

\bibitem[{{Zolotukhin} {et~al.}(2017){Zolotukhin}, {Bachetti}, {Sartore},
  {Chilingarian}, \& {Webb}}]{zolotukhin17}
{Zolotukhin}, I.~Y., {Bachetti}, M., {Sartore}, N., {Chilingarian}, I.~V., \&
  {Webb}, N.~A. 2017, \apj, 839, 125

\end{thebibliography}

\end{document}